\documentclass[11pt,a4paper]{article}
\pdfoutput=1
\usepackage{jheppub}
\usepackage[utf8]{inputenc}

\usepackage[T1]{fontenc}

\usepackage{amsmath}
\usepackage{amsfonts}
\usepackage{amssymb}
\usepackage{amsthm}
\usepackage{mathtools}
\usepackage{mathrsfs}

\usepackage{physics}
\usepackage{bbm}
\usepackage{dsfont}
\usepackage{ytableau}
\ytableausetup{smalltableaux}

\usepackage{longtable}
\usepackage{stackengine}
\usepackage{graphicx}
\graphicspath{{images/}}

\usepackage[normalem]{ulem}

\usepackage{hyperref}

\begin{document}
\title{\boldmath Monodromy line defects on the fuzzy sphere}

\author[a]{Jiangyuan Qian}
\affiliation[a]{Department of Particle Physics and Astrophysics, Weizmann Institute of Science, Rehovot 7610001, Israel}
\emailAdd{jiangyuan.2718@gmail.com}

\abstract{
Monodromy line defects are of interest in the defect bootstrap program and are dual to electrically charged objects in dual gauge field descriptions. We investigate the Ising monodromy dCFT using a fuzzy sphere regularization, by imposing the twisted boundary condition in the UV. We obtain defect scaling dimensions $\hat \Delta$ and ratios of defect operator expansion (DOE) coefficients using this method. We extend this machinery to the continuous family of $O(2)$ monodromy defects with twist angle $2\pi\nu$. In both cases, the lowest neutral defect scalar lies above dimension one, providing evidence that the monodromy defect reaches a stable dCFT without additional fine tuning.
}
\maketitle

\section{Introduction}
Conformal field theories (CFTs) are central to the study of quantum field theory; their enhanced symmetry constrains physical observables, allowing many quantities to be determined from a relatively small set of conformal data. In some cases these data can be computed using perturbative methods, such as the $\varepsilon-$expansion \cite{eps_exp} or the large-$N$ expansion \cite{Moshe:2003xn}. For some theories can be constrained or even be solved through the conformal bootstrap program \cite{CB,Simmons-Duffin:2016wlq,Chester:2020iyt}.

The fuzzy-sphere regularization provides a method for computing conformal data in $2+1$ dimensions \cite{Wei_23,Hu_23}, by constructing a model of interacting electrons restricted to the lowest Landau level (LLL) of a two sphere with $U(1)$ magnetic flux $\int da = 4\pi s$. If the model has a critical point in the universality class of the target CFT, its low energy spectrum realizes the CFT spectrum on $\mathbb S ^2 \times \mathbb R$. The state-operator correspondence then relates these states to local operators. This method has been applied to many strongly coupled CFTs, including Ising CFT, O(N) Wilson-Fisher CFTs, and Chern-Simons matter theories etc. \cite{Guo:2025odn,dey_26,Zhou_25,Dedushenko:2024nwi,Zhou:2023qfi,Feng:2026iii,Zhou:2025fer,Stergiou:2026cub,Han:2023lky}

A natural extension is to study theories whose symmetry is a subgroup of the $SO(D+1,1)$ conformal group, a special class being defect conformal field theories (dCFTs). Consider a codimension-p defect in D dimensional quantum field theory by wrapping a codimension-p object (for example on a hyperplane) in a way which is invariant under $SO(p)\times SO(D-p+1,1)$. The conformal symmetry allows us to map the theory to $\mathbb S^{D-1}\times \mathbb R$ with a defect on a $S^{D-p-1} \in S^{D-1}$; the $SO(D)$ spatial rotation symmetry on $\mathbb S^{D-1}$ is broken to the subgroup $SO(p)\times SO(D-p)$. In this work we focus on line defects in 3d i.e. $p=2,D=3$. There are new classes of operators $\hat O(x_\parallel)$ that live on the defect, the conformal data are the bulk data $\Delta_i, C_{ijk}$ (dimensions and 3-point functions of bulk operators), the defect data $\hat \Delta_i , f_{ijk}$ (dimensions and 3-point functions of defect operators), and the bulk-to-defect operator expansion coefficients $a_O,b_{O\hat O}$. From the state-operator correspondence, one may map the energy eigenstates on $\mathbb S^2 \times \mathbb R$ (in the presence of a defect at a point and an anti-defect at the antipodal point) to the defect spectrum, with $E_n-E_0=\frac{\hat \Delta_i}{R}$, where $R$ is the size of the sphere, and $\hat \Delta_i$ is the scaling dimension of defect operators.

Defect CFTs can be studied using the fuzzy-sphere regularization. Under radial quantization, a flat 1-dimensional defect can be constructed by point-like objects on the spatial $\mathbb S^2$, for instance, a $\mathbb Z_2$ breaking line defect can be constructed by adding point-like terms to the Hamiltonian $H_{d} =h n^z(\Omega_N)+hn^z(\Omega_S)$ on the Ising model, where $\Omega_{N,S}$ are the north and south poles of $\mathbb S^2$. This breaks the $\mathbb Z_2$ global symmetry, and breaks the $SO(3)$ rotation symmetry to $SO(2)$. $\mathbb Z_2$ breaking defects, $O(N)$ pinning defects etc. have been studied extensively on the fuzzy sphere \cite{Yang:2026qtk,Sarma:2026lte,Liu:2026dqc}. Also, large spin expansion \cite{Lemos:2017vnx}, g-function \cite{Zhou:2023fqu,Barkeshli:2025cjs,Cuomo:2021kfm} are used to study these line defects.

In this work, we focus on monodromy defects \cite{Billo:2013jda}, which are implemented by imposing twisted boundary conditions for the operators charged under a global symmetry $G$, i.e. $\Phi(\theta,\phi+2\pi)= g\Phi(\theta,\phi)$. One example is the Ising monodromy defect, for which $\Phi$ is the fundamental field and $g=-1$; this defect is dual to the $\mathbb Z_2$ Wilson lines in the dual gauge description \cite{Caselle:1996ii}. We also study the monodromy defects in the $O(2)$ Wilson-Fisher (WF) theory, where the fundamental field $\phi$ has monodromy $e^{-2\pi i \nu}$. With charge conjugation identifying $\nu \rightarrow-\nu$, there is a continuous family of theories with $\nu=(0,\frac{1}{2}]$. Both examples are of interest in the bootstrap program. For the Ising defect \cite{Gaiotto:2013nva} uses the crossing equation for a defect four-point function to determine the scaling dimension of the $p^0$ scalar, given the $\hat S ,\hat \psi_{1/2}$ scaling dimension data. For $O(2)$ defects, \cite{Gimenez-Grau:2022czc} shows agreement of $\epsilon$-expansion results and conformal bootstrap bounds, however, fewer non-perturbative data are available in this case, neither Monte Carlo data nor bootstrap island is known. This makes the fuzzy-sphere approach particularly useful as an independent non-perturbative probe. Monodromy defects have also been studied in  $SO(2N)$ theories \cite{Kravchuk:2025evf} and in the Wess-Zumino model \cite{GimenezGrau:2021wiv}.

The paper is organized as follows, in sections 2 and 3 we review conformal line defects and monodromy defects. In section 4 we will construct a fuzzy-sphere model for the Ising monodromy defects by imposing twisted boundary conditions to the monopole harmonics. Using exact diagonalization (ED) and density matrix renormalization group (DMRG), we compute several dCFT conformal data $\hat \Delta_i$ and ratios of DOE coefficients. In section 5 we will apply the same machinery to the $O(2)$ Wilson-Fisher theory and study a continuous family of monodromy defects with twist angle $2\pi\nu$.

\section{Conformal line defects in three dimensions}
A conformal line defect in 3d breaks the conformal symmetry $SO(4,1)$ to $SO(2,1)\times SO(2)$. The factor $SO(2,1)$ is the conformal group acting along the defect line, generated by $D,P,K$, where $D$ generates an overall dilatation $x^\mu \rightarrow\lambda x^\mu$, $P$ generates translation along the defect, and $K$ generates the special conformal transformation. By introducing inversion $I: x^\mu\rightarrow x^\mu/ x^2$, we can write $K= I^{-1} P I$. In the examples we consider, there is a parity symmetry $\mathcal P_\parallel$ that sends $x_\parallel\rightarrow-x_\parallel$. The transverse rotation group $SO(2)$ is generated by $L_z$, which commutes with the conformal generators, and the continuous symmetry is then denoted by $SO(2,1)^+\times SO(2)_r$.

A representation of the conformal algebra is specified by a primary with $|\Delta ,l_z\rangle$, satisfying $K|\Delta,l_z\rangle =0$, and its descendants have the form $P^k|\Delta,l_z\rangle$, with quantum numbers $|\Delta+k,l_z\rangle$. The parity operator $\mathcal P_\parallel$ anti-commutes with the translation generator $P$, thus the conformal tower is graded by $\mathcal P_\parallel$, i.e. $\hat {\mathcal P}_\parallel P^k|\Delta,l_z\rangle=\mathcal P_0 (-1)^kP^k|\Delta,l_z\rangle$, where $\mathcal P_0$ is the parity eigenvalue of the primary.

As the theory contains defect operators $\hat{O}(x_\parallel)$ and bulk operators $O(x_\parallel,x_\perp)$, we may consider different types of correlation functions. The non-identity defect one-point function $\langle\hat{O}(x_\parallel)\rangle_{\mathbb R^3}$ vanishes according to the $P$ and $D$ symmetries, but a bulk operator can have a non-zero one-point function of the form (in flat space):
\begin{equation}
    \langle O(x_\parallel,x_\perp)\rangle_{\mathbb R^3} = \frac{a_O}{|x_\perp |^{\Delta_O}}.
\end{equation}
The bulk-defect two point function is also constrained, for scalar defect operators it is given by \cite{Billo:2016cpy, Gadde:2016fbj}:
\begin{equation}
    \langle O(x_\parallel,x_\perp)\hat O(y_\parallel)\rangle_{\mathbb R^3}=\frac{b_{O\hat O}}{[(x_\parallel-y_\parallel)^2 + x_\perp^2]^{\Delta_{\hat  O}}|x_\perp|^{\Delta_O-\Delta_{\hat  O}} }.
\end{equation}
The two-point function of bulk operators depends on two cross-ratios $\frac{(x-y)^2}{4|x^\perp||y^\perp|},\frac{x^\perp\cdot y^\perp}{|x^\perp||y^\perp|}$. In a dCFT, a parity-even bulk operator admits defect operator expansion (DOE) of the form:
\begin{equation}
    O(x_\parallel,x_\perp) =  \frac{a_O}{|x_\perp|^{\Delta_O}}+ \sum _i\sum_{k=0}^{\infty} \frac{b_{O\hat O _i}^{(k)}}{|x_\perp|^{\Delta_O-\Delta_{\hat O_i}-2k}}\hat  P^{2k}\hat O_i(x_\parallel).
\end{equation}
Here we are ignoring parity-odd defect operators which can appear in less singular terms.
The coefficients $b_{O\hat O _i}^{(k)}$ within a $SO(2,1)$ conformal family are not independent, and the $\hat O_i$ are parity even defect primaries. One may insert a defect operator on both sides of the DOE and match the coefficients of $\frac{x_\perp^2}{(x_\parallel-y_\parallel)^2}$ on both sides. The DOE is specified by $a_O$  for each bulk operator and $b_{O\hat O}$ for each bulk-defect operator pair.

The state-operator correspondence also applies to dCFTs. In radial quantization, the past infinity is a point that lies on the defect, thus the Hilbert space states are generated by the defect operator $\hat O_i(0)$ at the origin of $\mathbb R^3$. In this basis, bulk insertions create superpositions of dilatation eigenstates $\hat O_i(0)$, as determined by the DOE.

It is useful to present the conformal mapping from flat space to a sphere. With the radial coordinate $r=e^\tau$, the metric is:
\begin{equation}
    ds^2 = e^{2\tau }(d\tau^2 +d\Omega^2_2).
\end{equation}
A conformal transformation $g\rightarrow e^{2\Omega} g$ where $\Omega = -\tau+ \log (R) $ maps a flat space to a cylinder $\mathbb S^2 \times \mathbb R$ of radius $R$, and correlators transform as:
\begin{equation}
    \langle \phi(x)...\rangle_{\mathbb S^2\times\mathbb R} = \left(\frac{e^{\tau}}{R}\right)^\Delta \langle \phi(x)...\rangle_{\mathbb R^3}\label{eq: conformal mapping}.
\end{equation}

For a conformal defect there is a Ward identity associated with the translation symmetry broken by the defect:
\begin{equation}
    \partial_\mu T^{\mu i }(x_\parallel,x_\perp) = \sqrt{C_D} \hat D^i(x_\parallel) \delta^{(p)}(x_\perp),
\end{equation}
with the normalization chosen such that $\hat D^i$ has the canonically normalized two-point function. From conformal symmetry one can show that $\hat D^i$ transforms as a vector under $SO(2)_r$ rotation and has scaling dimension two for a line defect in three dimensions. It is called the displacement operator and generates the transverse displacement $\xi^i(x_\parallel)$ of the defect line.

\section{Monodromy line defects and their fixed points}

A monodromy defect is a codimension-2 defect, around which local operators charged under the global symmetry group $G$ have a non-trivial symmetry action when cycling the monodromy defect, i.e.
\begin{equation}
    \Phi(r,\phi+2\pi,x_\parallel) = D(g) \Phi(r,\phi,x_\parallel), \qquad g \in G.
\end{equation}
In the Ising example, $G=\mathbb Z_2, g=-1$, a $\mathbb Z_2$-odd operator of dimension $\Delta$ has DOE:
\begin{equation}
    O(r,\phi,x_\parallel) = \sum_{i,n} e^{-i(n+\frac{1}{2})\phi}\frac{b_{O\hat O _i}}{r^{\Delta-\hat\Delta_i}} \hat O_{i,n+\frac{1}{2}}(x_\parallel), \qquad n\in \mathbb Z,
\end{equation}
where $\hat{O}$ has spin $n+\frac{1}{2}$ and scaling dimension $\hat \Delta$. In particular, the bulk spin operator $\sigma$ has DOE:
\begin{equation}
    \sigma(r,\phi,x_\parallel) = \frac{b_{\sigma\hat \psi_{1/2}}}{r^{\Delta_\sigma- \Delta_{\hat\psi_{1/2}}}}\left(e^{-i\phi /2}\hat \psi_{1/2}(x_\parallel)  +e^{i\phi /2}\hat{\bar\psi}_{1/2}(x_\parallel)   \right)+...,
\end{equation}
and the $\mathbb Z_2$-even operator $\epsilon$ has the expansion:
\begin{equation}
    \epsilon(r,\phi,x_\parallel) = \frac{a_\epsilon}{r^{\Delta_{\epsilon}}}+\frac{b_{\epsilon \hat S}}{r^{\Delta_\epsilon-\Delta_{\hat S}}}\hat S(x_\parallel)+\frac{b_{\epsilon\hat D}}{r^{\Delta_\epsilon- 2}}\left(e^{-i\phi }\hat D(x_\parallel)  +e^{i\phi }\hat{\bar D}(x_\parallel)  \right)+... .
\end{equation}
The discrete rotation symmetry $R_{yz}(\pi)$ preserves the defect geometry. Its action on the $\sigma$ operator $\sigma(r,\phi,x_\parallel)\rightarrow\sigma(r,-\phi,-x_\parallel)$, also preserves the monodromy condition.

We will also study a continuous family of $O(2)$ defects, a charge $Q$ operator under global $O(2)$ symmetry has monodromy condition $O_Q(r,\phi+2\pi,x_\parallel)=e^{-2\pi i \nu Q}O_Q(r,\phi,x_\parallel)$. The DOE reads:
\begin{equation}
    O_Q(r,\phi,x_\parallel) =\sum_{n,i} e^{-i(Q\nu+n)\phi} \frac{b_{O\hat O_i}}{r^{\Delta-\hat \Delta_i}} \hat O_{Q\nu +n,i}(x_\parallel\textbf{}).
\end{equation}
For example, the fundamental field operator $\phi$ has DOE:
\begin{equation}
    \phi(r,\phi,x_\parallel) = \frac{b_{\phi\hat \Psi_\nu
    }}{r^{\Delta_\phi- \Delta_{\hat \Psi_\nu}}}e^{-i\nu\phi} \hat\Psi_{\nu}(x_\parallel) +....
\end{equation}
Charge conjugation symmetry is preserved only at $\nu =0,1/2$, for other values, the change of variables $\phi = \tilde\phi^*$ reverses the twist angle from $2\pi \nu $ to $-2\pi\nu$. Identifying the twist angles related by charge conjugation, we can restrict the non-trivial twist to $\nu \in(0,\frac{1}{2}]$. The neutral bulk scalar $S$ has DOE:
\begin{equation}
    S(r,\phi,x_\parallel) = \frac{a_S}{r^{\Delta_{S}}}+\frac{b_{S \hat S}}{r^{\Delta_S-\Delta_{\hat S}}}\hat S(x_\parallel)+\frac{b_{S\hat D}}{r^{\Delta_S- 2}}\left(e^{-i\phi }\hat D (x_\parallel) +e^{i\phi }\hat {\bar D} (x_\parallel)  \right)+....
\end{equation}
For a generic value of $\nu$, the $R_{yz}$ and $C$ symmetry are individually broken, the combination $ R_{yz}C$ is a preserved $\mathbb Z_2$ symmetry of the theory, denoted by $\mathbb Z_{2,RC}$.

For a free (complex) scalar in $2+1$ dimensions, the bulk operator equation of motion gives the scaling dimensions $(\Delta_{\Psi_s}-\frac{1}{2})^2 -s^2 =0$. For the Wilson-Fisher fixed point, one may use the $\varepsilon$-expansion around $3+1$ dimensions \cite{Soderberg:2017oaa,Giombi:2021uae, Yamaguchi_2016,Rychkov:2015naa} to compute the scaling dimensions of spin-$s$ defect operators:
\begin{equation}
    \Delta_{\Psi_s} = 1+|s| -\frac{\varepsilon}{2}+\frac{1}{5}\frac{\nu(\nu-1)}{|s|}\varepsilon +O(\varepsilon^2),\label{eq: O2 charge mode}
\end{equation}
and of defect scalar primary $\hat S$:
\begin{equation}
    \Delta_{\hat S} = 2+2\nu -\varepsilon +\varepsilon\left(\frac{4}{5(1+2\nu)}+\frac{2(\nu-1)}{5}\right)+O(\epsilon^2).\label{eq: O2 neutral mode}
\end{equation}

\begin{figure}[t]
    \centering
    \begin{minipage}[t]{0.48\linewidth}
    \stackinset{l}{2mm}{t}{2mm}{\textbf{a.}}{
        \includegraphics[width=\linewidth]{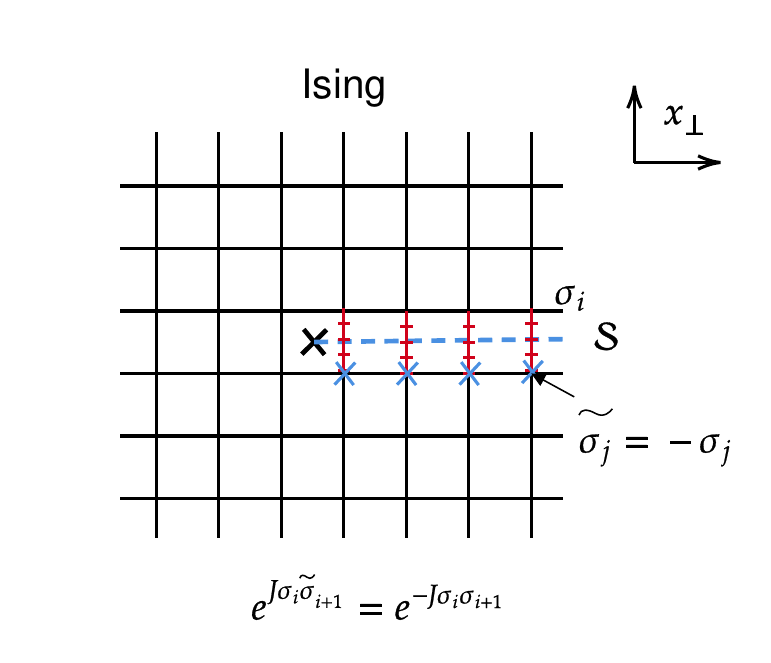}
        }
    \end{minipage}
     \begin{minipage}[t]{0.48\linewidth}
    \stackinset{l}{2mm}{t}{2mm}{\textbf{b.}}{
        \includegraphics[width=\linewidth]{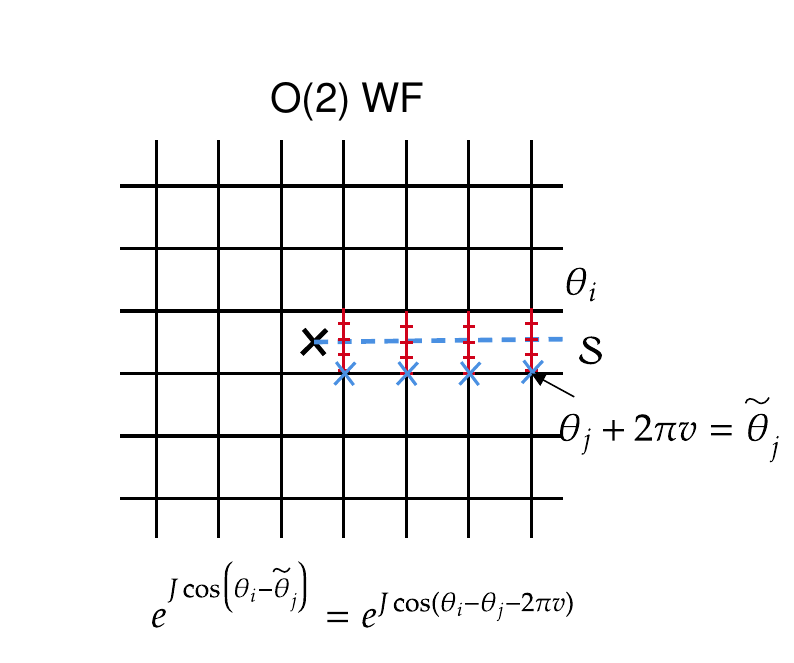}
        }
    \end{minipage}

    \caption{Lattice implementations of (a) Ising monodromy line defect (b) $O(2)$ monodromy line defect, each panel shows a cross-section in $x_\perp$ plane perpendicular to the defect. }
    \label{fig:lattice_realization}
\end{figure}

These CFTs can be constructed on a lattice by adding a topological surface $\mathcal S$ to the critical bulk model, which ends on the line defect and appear as 'branch-cuts' in the 2-plane transverse to the defect. For the Ising model one may introduce a second branch spin $\tilde \sigma$ with identification $\tilde\sigma=-\sigma$ on one side of the surface $\mathcal S$, see figure \ref{fig:lattice_realization}. A change of variables then gives frustrated couplings $J\rightarrow-J$ on links crossing the topological surface $\mathcal S$. The position of the topological surface $\mathcal S$ is not physical, since a deformation $\mathcal S \rightarrow \mathcal S'$ can be done by flipping the spins in $\Sigma$ with $\partial \Sigma =\mathcal S -\mathcal S'$. In the compact $O(2)$ model, the two branches are identified by $\tilde\theta_j = \theta_j +2\pi\nu$, the frustrated links contribute a Boltzmann factor:
\begin{equation}
    \prod_{\langle ij\rangle\text{ cross }\mathcal S}e^{J\cos(\theta_i-\theta_j)}\rightarrow\prod_{\langle ij\rangle\text{ cross }\mathcal S}e^{J\cos(\theta_i-\theta_j-2\pi \nu)}.\label{eq: frustrated_link}
\end{equation}
For continuum $O(2)$ model, one could turn on a background gauge field $A=\nu d\theta$ for the $O(2)$ global symmetry, this absorbs the twisted phase through a gauge transformation and makes $\phi$ untwisted, and this description is manifestly $O(2)$ rotational-symmetric.

\section{Ising monodromy line defects on the fuzzy sphere}
\subsection{Model construction}
In the fuzzy sphere regularization of the 3d Ising CFT, one takes two fermion flavors $\Psi=(\psi_\uparrow,\psi_\downarrow)^T$ on $\mathbb S^2$, coupled to a magnetic monopole background with flux $\int dA = 4\pi q$, $q=1/2,1,3/2,...$. After projecting to the lowest Landau level, the fermion annihilation field has the mode expansion $\hat \psi_a=\frac{1}{R}\sum_{m=-q}^q Y_{q,m}^{(q)}(\Omega) \psi_{a,m}$, each flavor has $N=2q+1$ orbitals, $R=\sqrt{N}$ is the size of the sphere, and the Ising model is realized by the quantum Hamiltonian:
\begin{equation}
    \begin{split}
        H_{\text{int}}=&\frac{1}{2}\int d^{2}r ::n^{0}\left(\vec{r}\right)\left(V_0-V_1\nabla^{2}\right)n^{0}\left(\vec{r}\right)::\\-&\frac{1}{2}\int d^{2}r ::n^{z}\left(\vec{r}\right)\left(V_0-V_1\nabla^{2}\right)n^{z}\left(\vec{r}\right)::\\
        -& h_x \int d^2 r n^x(\vec r),\label{eq: Hamiltonian_Ising_continuum}
    \end{split}
\end{equation}
where $n^0= \Psi^\dagger \Psi$, $n^a= \Psi^\dagger \sigma^a \Psi$, $::$ denote normal ordering, $\sigma^a$ are Pauli matrices acting on the fermion flavors. In addition to the spatial $SO(3)_r$ symmetry, this model has a flavor $\mathbb Z_{2\mathcal I}$ symmetry, whose action is:
\begin{equation}
    \mathcal I : \Psi\rightarrow\sigma^x \Psi.
\end{equation}
When increasing $h_x$ the model undergoes a second-order phase transition from a ferromagnetic (FM) phase to a paramagnetic (PM) phase. At the critical point $h_{x,c}$ the low-energy spectrum exhibits emergent conformal symmetry, see \cite{Wei_23}.

Consider a line defect stretching along the $x^3$ direction. The conformal mapping from $\mathbb R^3$ to $\mathbb S^2\times \mathbb R$ maps the line to branch points at the north and south poles of the two-sphere. We introduce the monodromy defect by requiring that the global symmetry in the UV has the same twisted boundary condition as the IR i.e. 
\begin{equation}
     \Psi(\theta,\phi)=D(g) \Psi(\theta,\phi+2\pi),
\end{equation}
where $g=-1$, $D(g)=\sigma^x$. One can diagonalize $D(g)$ in a new basis, $\Psi=(\psi_\rightarrow,\psi_\leftarrow)$, where the $\psi_{\rightarrow/\leftarrow}$ are the $\pm1$ eigenstates of $\sigma^x$. Then the monodromy reads:
\begin{equation}
\begin{split}
     \psi_\rightarrow(\theta,\phi+2\pi) =  +\psi_\rightarrow(\theta,\phi),\\
     \psi_\leftarrow(\theta,\phi+2\pi) =  -\psi_\leftarrow(\theta,\phi).
\end{split}
\end{equation}
We solve the monopole-harmonics \cite{Wu:1976ge} for the antiperiodic fermion $\psi_\leftarrow$, the derivation is in appendix \ref{app: solve_twist_modes},
\begin{equation}
\begin{split}
    Y_{q,m}^{(q)}\left(x,\phi\right)&=\mathcal{N}_{q,m}(1-x)^{\frac{q+m}{2}}(x+1)^{\frac{q-m}{2}}e^{i(q+m)\phi}.
\end{split}
\end{equation}
The monopole harmonics have the same analytic form for $\psi_\rightarrow$ and $\psi_\leftarrow$, with different allowed values of $m$:
\begin{equation}
    \begin{split}
        \psi_\rightarrow(\Omega)&= \frac{1}{R}\sum_{m=-q}^{q}\psi_{\rightarrow,m}Y^{(q)}_{q,m}(\Omega),\\
        \psi_\leftarrow(\Omega)&= \frac{1}{R}\sum_{m=-q+\frac{1}{2}}^{q-\frac{1}{2}}\psi_{\leftarrow,m}Y^{(q)}_{q,m}(\Omega).
    \end{split}
\end{equation}

We used the same form of Hamiltonian density as (\ref{eq: Hamiltonian_Ising_continuum}), inserting a completeness relation and doing the spatial integral gives (\ref{eq: Hamiltonian_Ising_expansion}). The UV model also has the discrete symmetry $R_{yz}(\pi)$\footnote{The transformation gives an extra phase $e^{2iq\phi}$ for fermions, this can be ignored if  the Hamiltonian is built from the density and its gradients.}, whose action on LLL modes is:
\begin{equation}
    R_{yz}\psi_m R_{yz}^{-1} = \psi_{-m}.
\end{equation}
Since the monodromy condition imposes a constraint $e^{2\pi i L_z}\circ \mathcal I=1$, we omit the $\mathbb{Z}_{2 \mathcal I}$ charge which can be read off from $L_z$. The states are labeled by quantum number $(L_z^{\mathcal R_{yz}})$. The $\mathbb Z_{2,yz}$ symmetry anti-commutes with $L_z$, so it is a quantum number in the $L_z=0$ sector, and identifies the spectrum at $L_z$ and $-L_z$. As the total orbital number of all the fermions is now odd, the particle-hole symmetry only exists in thermodynamic limit, where it may be identified with the $\mathcal P _\parallel$ symmetry.

We set $V_0=1$ as the interaction energy scale, the phase diagram as a function of $h_x$ is then shown in appendix \ref{app: phase transition}. Similar to the bulk model, we find a second-order phase transition from finite-size scaling of energy gaps, and we tune $V_1, h_x$ to the fixed point. More systematically, we define a cost function based on finite-size energy spectra, where for a relativistic critical system, the energy gaps scale as:
$$
E_n - E_0 = v\frac{\Delta}{R}
$$
for some constant $v$. Numerically, we fit the gaps by $E_n(N)-E_0(N)= \Delta E_n(\infty)+ \frac{v\Delta}{\sqrt{N}} $ from CFT scaling, and use an offset $\Delta E_n(\infty)$ to quantify the deviation from it:
\begin{equation}
    P(V_1,h_x)= \sum_{i=1}^{n_{max} } \frac{|\Delta E_i(\infty)|}{\langle\Delta E_i\rangle_N}\label{eq: fs_cost}.
\end{equation}
We use the first excited states in the $(0^+),(0^-),(1/2),(1), (3/2)$ sectors and the values $N=9,..,14$ in the function $P$. Minimizing $P$ gives the parameters:
\begin{equation}
    V_1 = 0.34396,\qquad h_x =0.00036.\label{eq:parameters}
\end{equation}

\begin{figure}[t]
    \centering
    \begin{minipage}[t]{0.45\linewidth}
    \stackinset{l}{2mm}{t}{2mm}{\textbf{a.}}{
        \includegraphics[width=\linewidth]{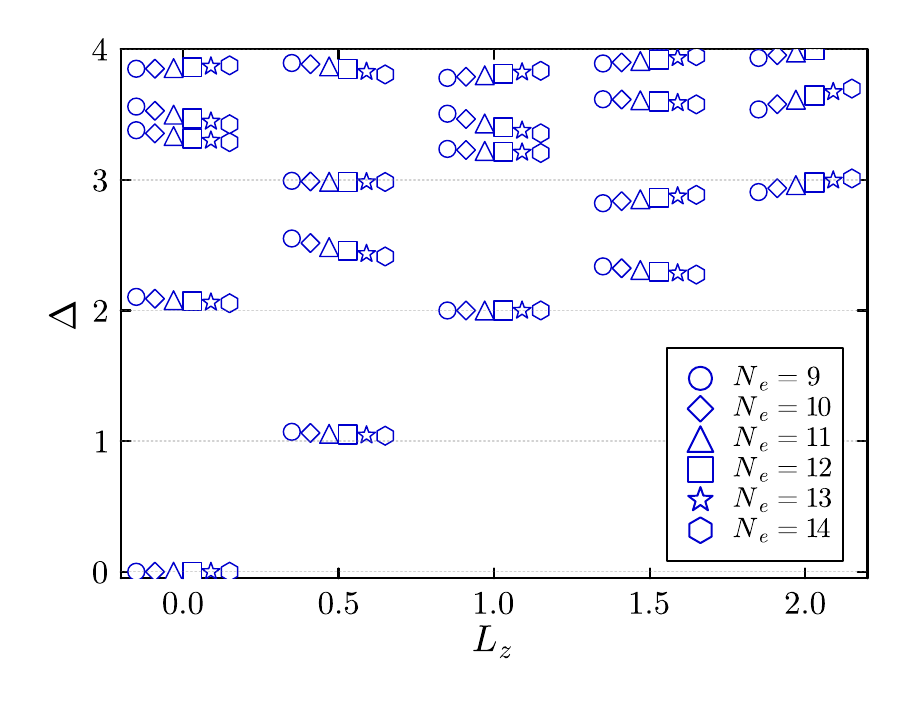}
        }
    \end{minipage}
     \begin{minipage}[t]{0.45\linewidth}
    \stackinset{l}{2mm}{t}{2mm}{\textbf{b.}}{
        \includegraphics[width=\linewidth]{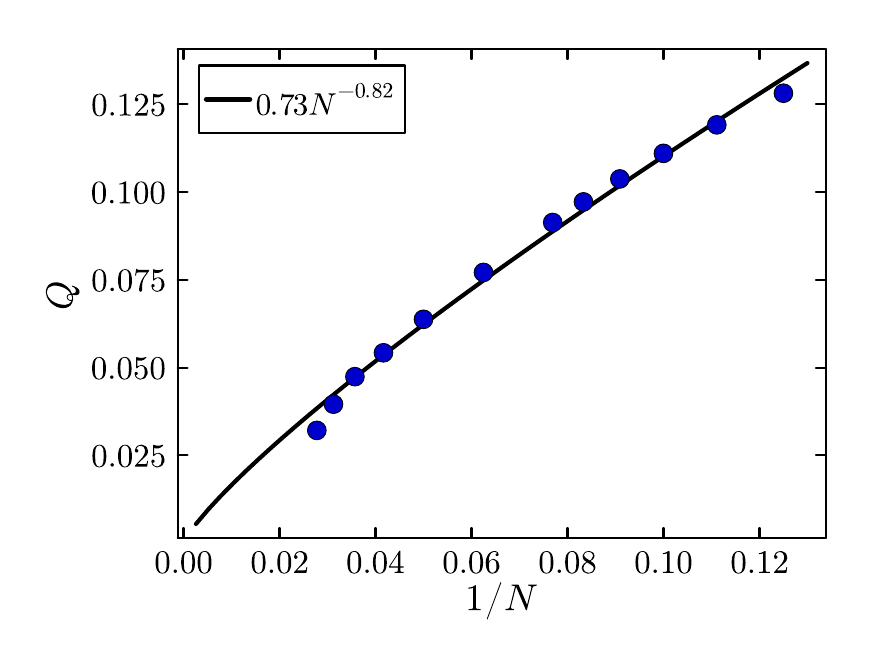}
        }
    \end{minipage}
    \begin{minipage}[t]{0.45\linewidth}
    \stackinset{l}{2mm}{t}{2mm}{\textbf{c.}}{
        \includegraphics[width=\linewidth]{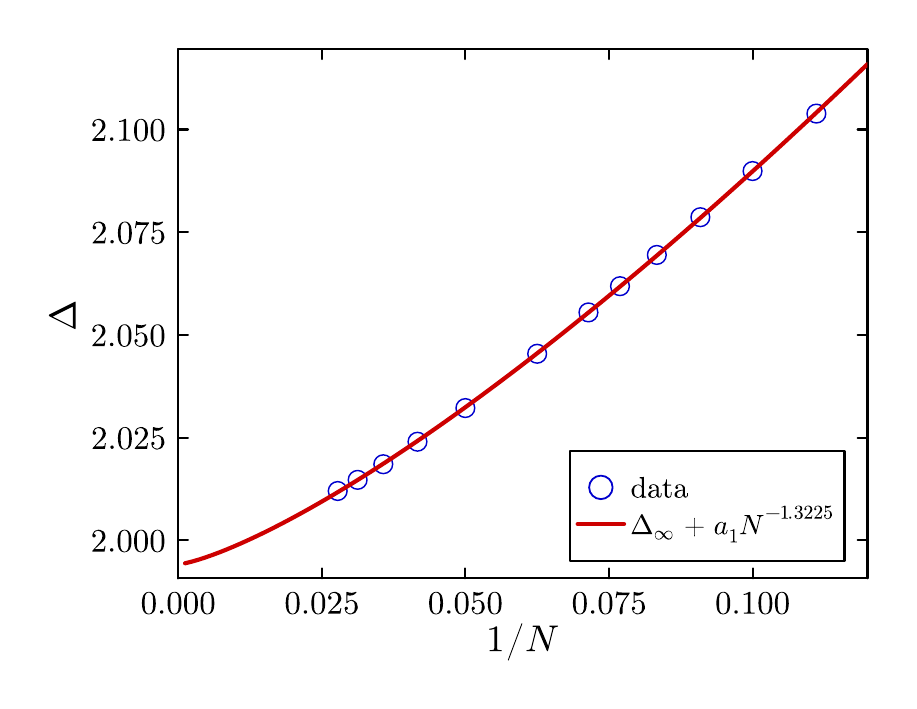}
        }
    \end{minipage}
    \begin{minipage}[t]{0.45\linewidth}
    \stackinset{l}{2mm}{t}{2mm}{\textbf{d.}}{
        \includegraphics[width=\linewidth]{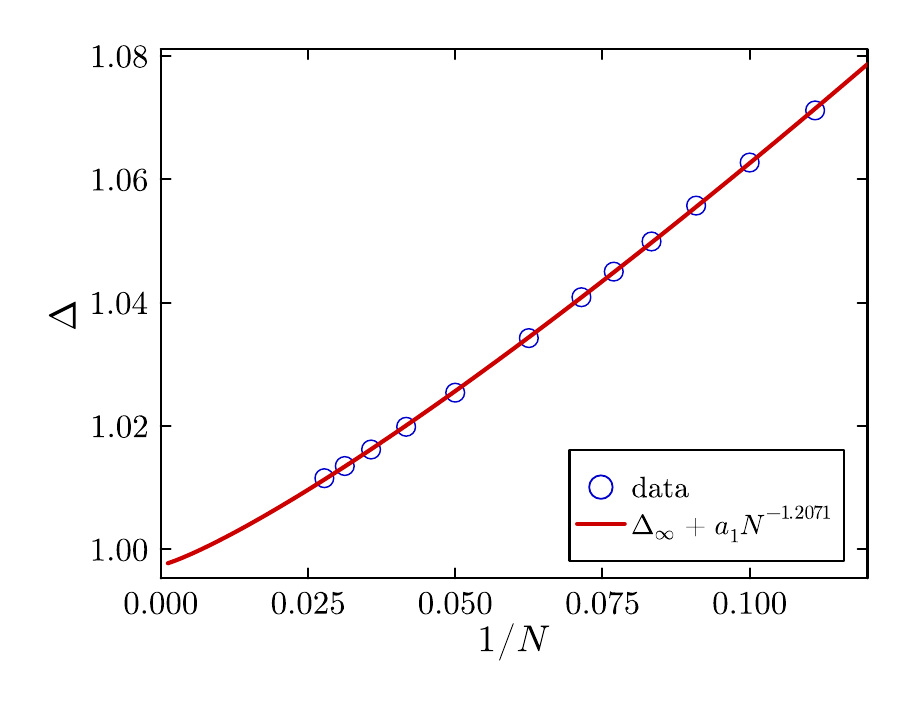}
        }
    \end{minipage}
    \begin{minipage}[t]{0.45\linewidth}
    \stackinset{l}{2mm}{t}{2mm}{\textbf{e.}}{
        \includegraphics[width=\linewidth]{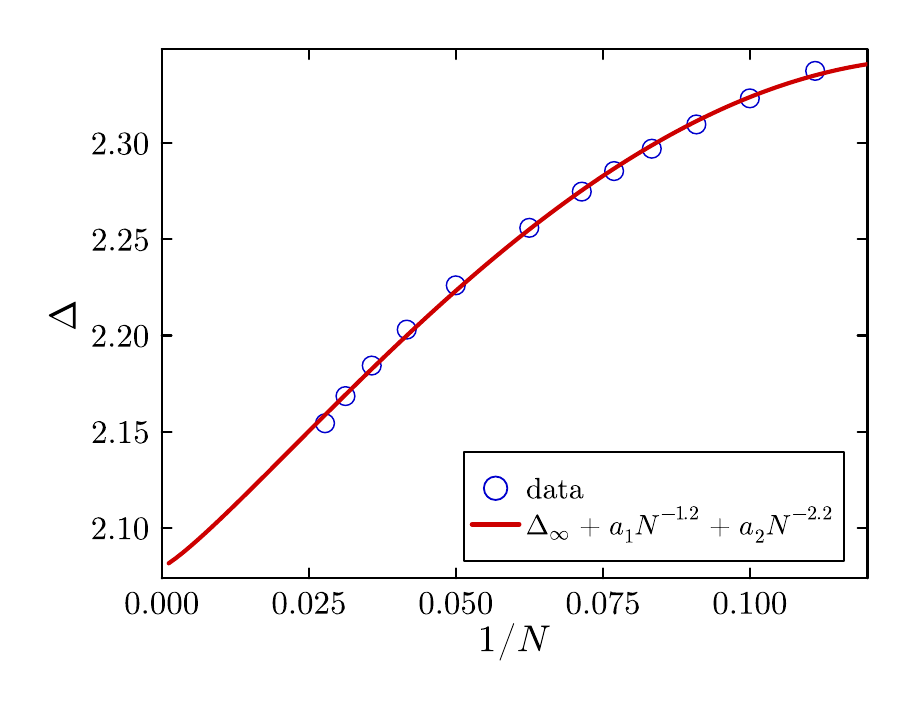}    
        }
    \end{minipage}
    \begin{minipage}[t]{0.45\linewidth}
    \stackinset{l}{2mm}{t}{2mm}{\textbf{f.}}{
        \includegraphics[width=\linewidth]{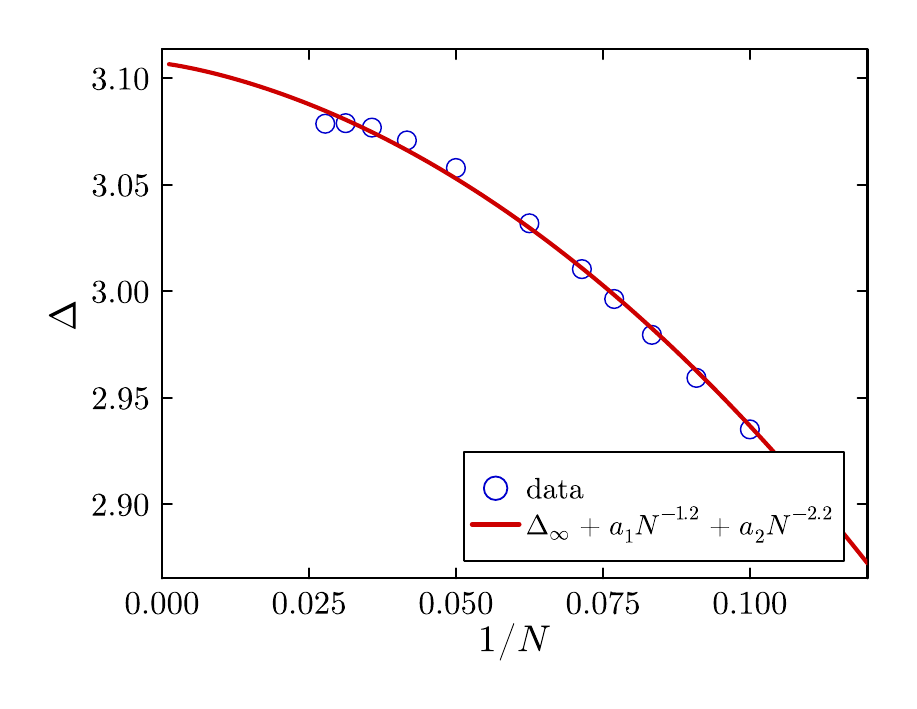}    
        }
    \end{minipage}

    \caption{(a) Conformal tower by ED at the tuned parameters \eqref{eq:parameters} and $N=9,...,14$, (b) the conformal cost for system size up to $N=36$. (c) Finite-size fits of scaling dimension $\Delta_{\hat S}$, (d) $\Delta_{\hat \psi_{1/2}}$, (e) $\Delta_{\hat \psi_{3/2}}$, (f) $\Delta_{t}$. }
    \label{fig: Del_fs_Ising}
\end{figure}

At these UV parameters, we obtain the conformal tower after setting $v$ such that $\Delta_D=2$, shown in figure \ref{fig: Del_fs_Ising}(a). To quantify the deviation of level spacing from dCFT, we define the conformal cost $Q(N)$ in appendix \ref{app: cost functions}. As shown in figure \ref{fig: Del_fs_Ising}(b), $Q(N)$ decreases as $N$ increases, showing evidence that the spectrum approaches the expected conformal structure in the thermodynamic limit.

\begin{table}[t]
  \setlength{\tabcolsep}{22pt}
   \renewcommand{\arraystretch}{1.0}
    \centering
    \begin{tabular}{cccccc}
    \hline \hline
       $\hat O$  &  $L_z$ & $\hat\Delta^{N=14}$& $\hat\Delta^{\text{fit}}$& $\hat\Delta^{N=36}$  & $\hat\Delta^{MC}$   \\
        \hline
       $\psi_{1/2}$ &$\frac{1}{2}$ & 1.04&1.00(1) & 1.01 & 0.9187(6)\\
        $S$ & $0^+$ &2.06& 1.99(2)& 2.01 & 2.27(1)\\
       $\psi_{3/2}$ &$\frac{3}{2}$ & 2.27&2.08(7)& 2.15 & 1.99(5)\\
       $D^i$ & $1$ & 2.0&2.0  & 2.0  &2.0 \\
       $t_+$ &$2$  & 3.01& 3.11(3)&3.08 &3.1(5)\\
       $p^0$ &$0^+$  &3.43 & -- & 3.23 &2.9(2)\\
        \hline\hline
    \end{tabular}
    \caption{Estimates of defect scaling dimensions in Ising monodromy dCFT, compared with MC results.}
    \label{tab:scaling_dim_primary}
\end{table}

\subsection{Spectrum}
At the critical point, the finite size effect on scaling dimensions due to irrelevant perturbations takes the form
\begin{equation}
    \Delta(N) \simeq \Delta (\infty) + a_1 \frac{1}{N^{\omega/2}}+ a_2 \frac{1}{N^{(\omega+2)/2}}+...,
\end{equation}
where $a_1$ denote the leading irrelevant perturbation correction. We fit the finite-size data using this form, as shown in figure \ref{fig: Del_fs_Ising}(c-f). The finite-size extrapolation of the lowest operators is shown in table \ref{tab:scaling_dim_primary}, having discrepancies of order ten percent with the Monte Carlo results. The lowest neutral scalar on the defect has $\Delta>1$, suggesting the monodromy condition is sufficient to reach a stable IR dCFT, without additional defect fine-tuning.

\subsection{DOE coefficients}
A UV operator can be expanded in local bulk CFT operators:
\begin{equation}
\begin{split}
     n^x(\Omega) &= c_I I + c_\epsilon \epsilon(\Omega) +...,\\
     n^z(\Omega) &=  c_\sigma \sigma(\Omega) +...\
\end{split}
\end{equation}
Using the conformal map in \eqref{eq: conformal mapping}, we obtain
\begin{equation}
\langle\epsilon(\Omega)...\rangle_{\mathbb S^2\times \mathbb R} = \frac{1}{R^{\Delta_\epsilon}}\langle\epsilon(x_\parallel,x_\perp)...\rangle_{\mathbb R^3}.
\end{equation}
The matrix elements of bulk operator between vacuum and a generic defect primary state reads:
\begin{equation}
\langle \hat I| n^x(\Omega)|\hat O\rangle = \frac{c_\epsilon}{R^{\Delta_\epsilon}} \langle \hat I|\epsilon(x_\parallel,x_\perp)|\hat O\rangle =  \frac{c_\epsilon}{R^{\Delta_\epsilon}} b_{\epsilon\hat O} \frac{1}{(x^2)^{ \Delta_{\hat O}} |x_\perp|^{\Delta_{\epsilon}- \Delta_{\hat O}}},
\end{equation}
Setting $\sqrt{x^2} = e^\tau =1$, and $|x_\perp| = e^\tau \sin\theta $, gives the matrix elements on $\mathbb S^2$:
\begin{equation}
\langle \hat I| n^x(\theta,\phi=0)|\hat O\rangle =   \frac{c_\epsilon}{R^{\Delta_\epsilon}}  \frac{b_{\epsilon\hat O}}{(\sin\theta)^{\Delta_\epsilon- \Delta_{\hat O}}}.\label{eq: DOE coefficients}
\end{equation}

\begin{figure}[t]
    \centering
    \begin{minipage}[t]{0.45\linewidth}
    \stackinset{l}{2mm}{t}{2mm}{\textbf{a.}}{
        \includegraphics[width=\linewidth]{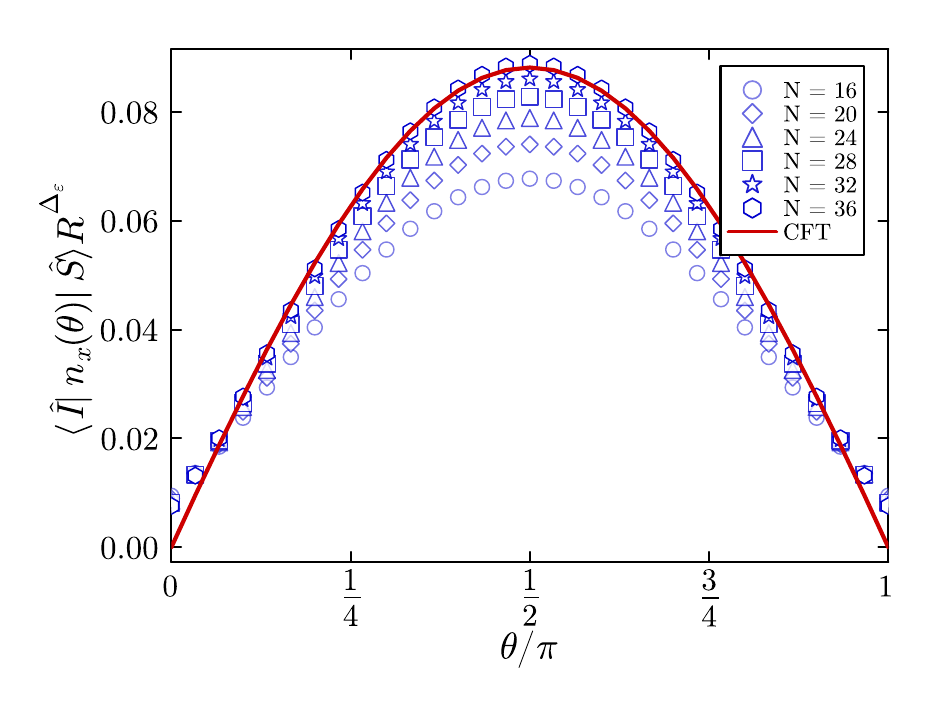}
        }
    \end{minipage}
     \begin{minipage}[t]{0.45\linewidth}
    \stackinset{l}{2mm}{t}{2mm}{\textbf{b.}}{
        \includegraphics[width=\linewidth]{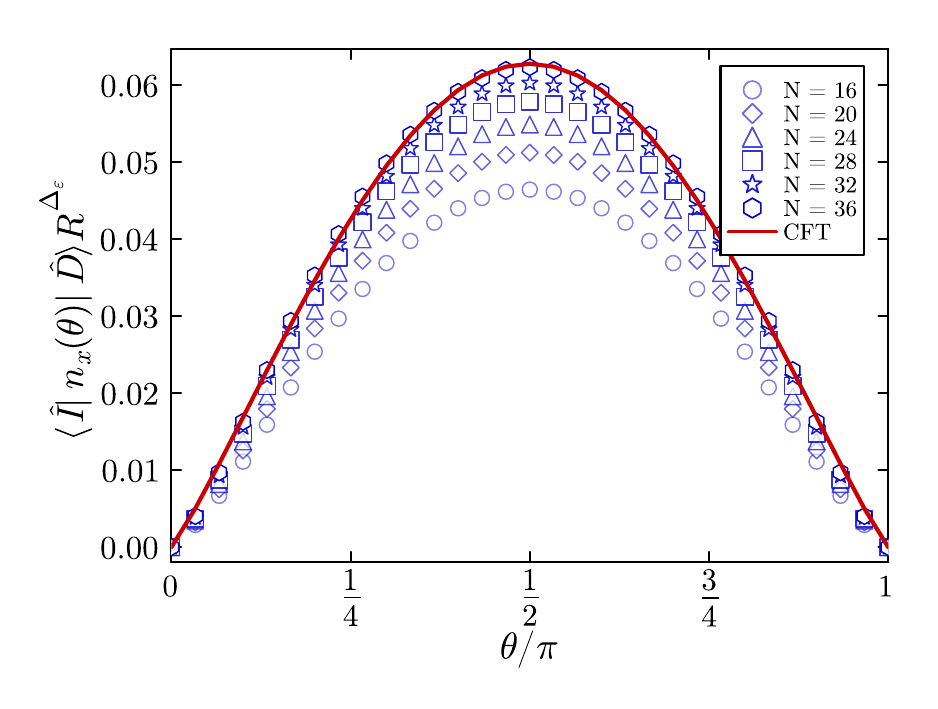}
        }
    \end{minipage}
    \begin{minipage}[t]{0.45\linewidth}
    \stackinset{l}{2mm}{t}{2mm}{\textbf{c.}}{
        \includegraphics[width=\linewidth]{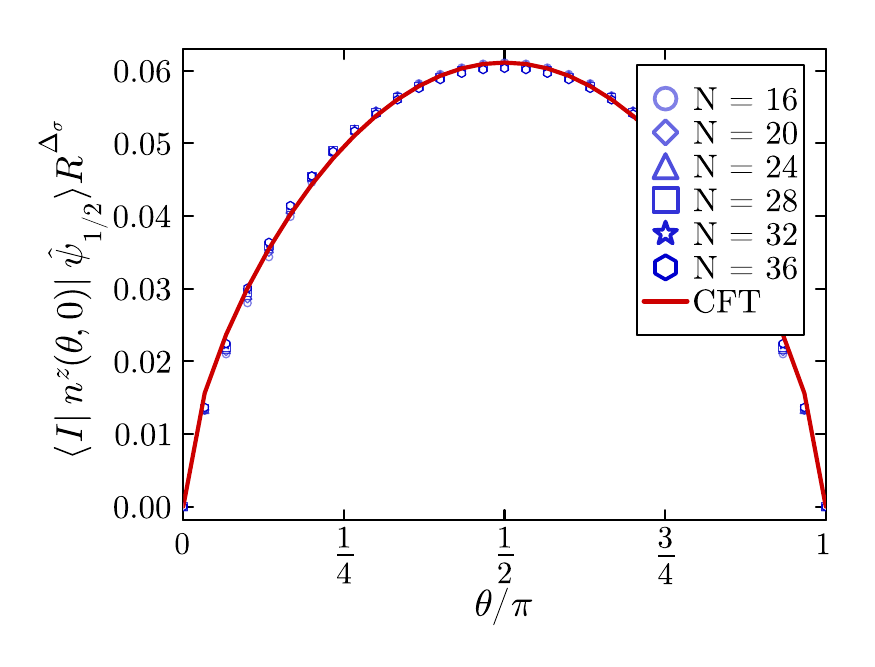}
        }
    \end{minipage}
    \begin{minipage}[t]{0.45\linewidth}
    \stackinset{l}{2mm}{t}{2mm}{\textbf{d.}}{
        \includegraphics[width=\linewidth]{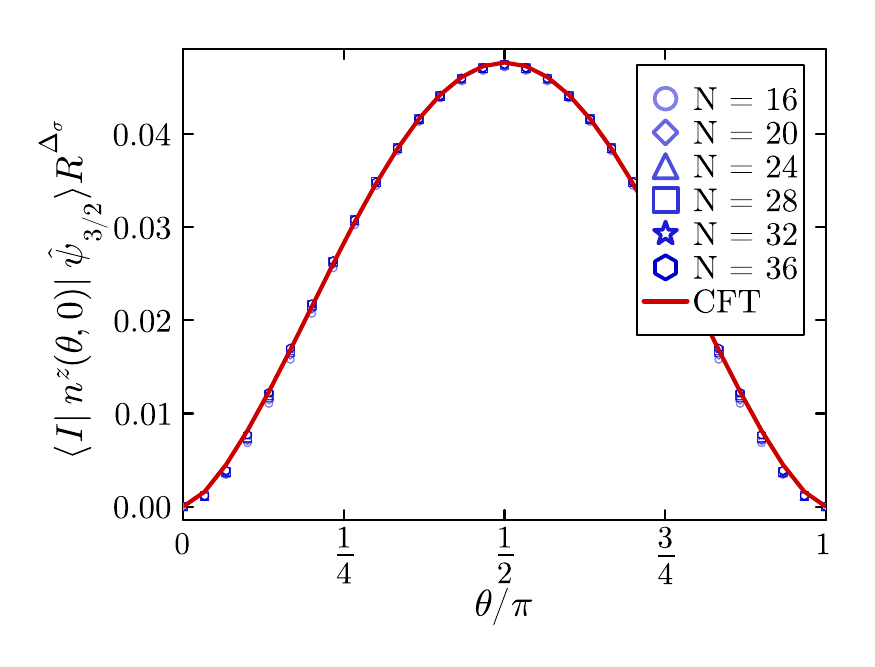}
        }
    \end{minipage}

    \caption{(a) Vacuum to primary matrix elements $\langle \hat I| n^x(\Omega)|\hat S\rangle$, (b)  $\langle \hat I| n^x(\Omega)|\hat D\rangle$, (c) $\langle \hat I| n^z(\Omega)|\hat \psi_{1/2}\rangle$, (d) $\langle \hat I| n^z(\Omega)|\hat \psi_{3/2}\rangle$. The CFT curves are fitted without fixing the bulk and defect scaling dimensions}
    \label{fig:one point function}
\end{figure}

\begin{figure}[t]
    \centering
    \begin{minipage}[t]{0.45\linewidth}
    \stackinset{l}{2mm}{t}{2mm}{\textbf{a.}}{
        \includegraphics[width=\linewidth]{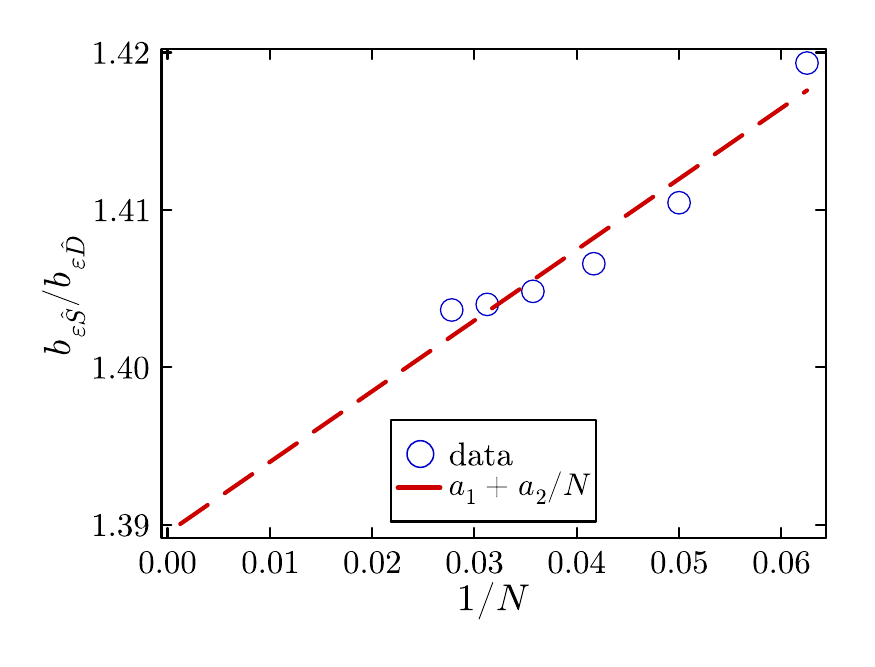}
        }
    \end{minipage}
     \begin{minipage}[t]{0.45\linewidth}
    \stackinset{l}{2mm}{t}{2mm}{\textbf{b.}}{
        \includegraphics[width=\linewidth]{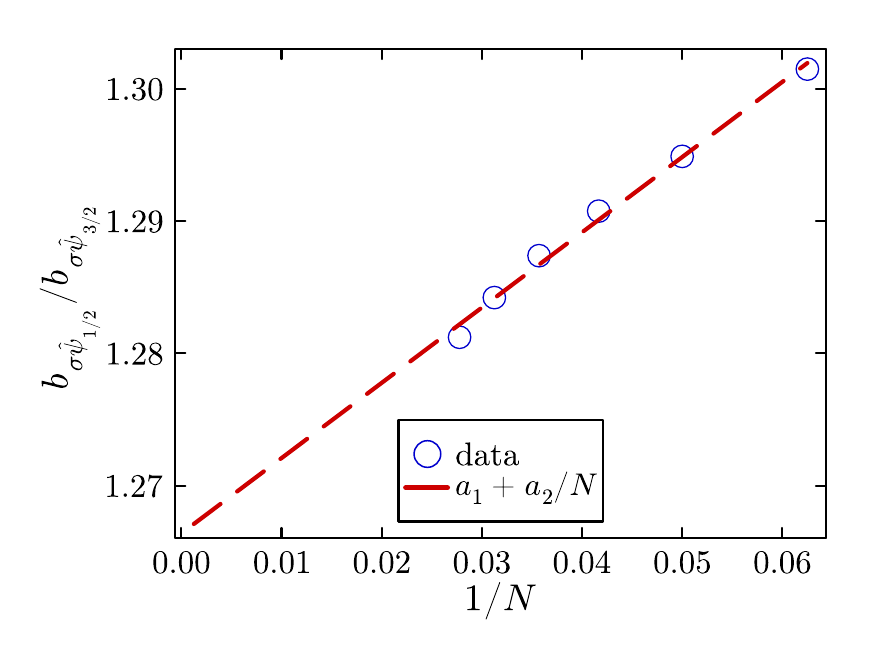}
        }
    \end{minipage}

    \caption{Finite-size extrapolations of ratios of DOE coefficients for the leading defect operators appearing in the expansion  of bulk operator (a) $\epsilon$ and (b) $\sigma$.}
    \label{fig: DOE_ratio}
\end{figure}

$n^z$ has non-zero vacuum-to-primary matrix element for half spin primaries, as shown in figures \ref{fig:one point function}(c-d). We extract $b_{\epsilon\hat S}/b_{\epsilon\hat D}$ and $b_{\sigma\hat \psi_{1/2}}/b_{\sigma\hat \psi_{3/2}}$ by fitting the data at $N=12,...,36$. The extrapolations are shown in figure \ref{fig: DOE_ratio}, and the results are listed in table \ref{tab: b_conformal data}.

\begin{table}[t]
    \setlength{\tabcolsep}{8pt}
   \renewcommand{\arraystretch}{1.2}
    \centering
    \begin{tabular}{c|c|c|c}
    \hline \hline
          $c_\epsilon b_{\epsilon\hat S}$ & $c_\epsilon b_{\epsilon\hat D}$  & $b_{\epsilon\hat S}/b_{\epsilon\hat D}$& $b_{\epsilon\hat S}/b_{\epsilon\hat D}$,MC \\
       \hline
         0.088(3)&0.063(2)&1.39(1)&--\\
        \hline \hline
          $c_\sigma b_{\sigma\hat \psi_{1/2}}$  &$c_\sigma b_{\sigma\hat \psi_{3/2}}$  & $b_{\sigma\hat \psi_{1/2}}/b_{\sigma\hat \psi_{3/2}}$& $b_{\sigma\hat \psi_{1/2}}/b_{\sigma\hat \psi_{3/2}}$,MC \\
       \hline
        0.0619(3)&0.0483(1)& 1.28(1)& 1.58(23)\\
        \hline \hline
    \end{tabular}
    \caption{DOE data extracted from vacuum-to-primary matrix elements, at tuned parameters and extrapolate using $N=12,16,...,36$.}
    \label{tab: b_conformal data}
\end{table}

\section{\texorpdfstring{O(2) WF monodromy lines on fuzzy sphere}{O2 WF monodromy lines on fuzzy sphere}}
\subsection{Model construction}
The O(2) Wilson--Fisher (WF) fixed point can be realized by a spin-1 model\footnote{Where the spin symmetry commutes with the spatial rotation group $SO(3)_r$ and act as a flavor symmetry.} on the fuzzy sphere \cite{dey_26}. We introduce a monodromy line defect by imposing twisted boundary condition to a fermion flavor of O(2) charge $\sigma$ (where the O(2) charge is the $S_z$ spin):
\begin{equation}
    \psi_{\sigma}^\dagger(\theta,\phi+2\pi) = e^{-2\pi i \nu \sigma}\psi_{\sigma}^\dagger(\theta,\phi),\qquad \sigma=-1,0,+1.
\end{equation}
The LLL modes of the three flavors have the same analytic form of the monopole harmonics, with $m$ chosen corresponding to the twisted boundary condition (see appendix \ref{app: solve_twist_modes}), the mode expansion reads:
\begin{equation}
    \begin{split}
        \psi_{1}(\Omega)&=\frac{1}{R}\sum_{m=-q+\nu}^{q-1+\nu} \psi_{1,m} Y^{(q)}_{q,m}(\Omega),\\
        \psi_{0}(\Omega)&=\frac{1}{R}\sum_{m=-q}^{q} \psi_{0,m}Y^{(q)}_{q,m}(\Omega),\\
        \psi_{-1}(\Omega)&=\frac{1}{R}\sum_{m=-q+1-\nu}^{q-\nu}\psi_{-1,m} Y^{(q)}_{q,m}(\Omega).\label{eq: mode_exp_O2}
    \end{split}
\end{equation}
Our choice of mode function is consistent with the free field branch it lies in; considering a free complex scalar field with monodromy angle $2\pi\nu$, the scaling dimension of defect operators with spin reads:
\begin{equation}
    \begin{split}
        \Delta_{\Psi_\nu}&=\frac{1}{2}\pm |\nu|,\\
        \Delta_{\Psi_{\nu-1}}&=\frac{1}{2}\pm |1-\nu|.
    \end{split}
\end{equation}
These defect operators arise from DOE of the bulk operator $\phi$ approaching the defect:
\begin{equation}
    \begin{split}
        \phi&\sim \Psi_\nu x_{\perp}^{\pm|\nu|}+\Psi_{\nu-1} x_{\perp}^{\pm|\nu-1|},\\
        \phi&\sim\Psi_\nu(1-x)^{\pm|\nu|/2}+\Psi_{\nu-1}(1-x)^{\pm|1-\nu|/2}, \\
    \end{split}
\end{equation}
where $|x_\perp|=r\sin\theta\sim(1-x)^{1/2}$, $x=\cos\theta$. Had we chosen the minus branch, $\phi$ would diverge near the defect. The defect conformal algebra $[K,P]=2D$ gives a unitarity bound $\Delta\geq0$, which means the diverging branch is not physical for the $ \Psi_{\nu-1}$ operator.\footnote{The bars on the defect operators are defined by: $\bar\Psi_s= (\Psi_{-s})^\dagger$.}

This observation relates to the choice of LLL modes among the normalizable modes, the singular mode contribution is:
\begin{equation}
    \begin{split}
        n^-(x,0) \sim \psi_{-1}^\dagger\psi_{0}+\psi_{0}^\dagger\psi_{1}\sim a_1 (1-x)^{(\nu-1)/2}+a_2 (1-x)^{-\nu/2}.
    \end{split}
\end{equation}
As $n^{-}\sim \phi^*$, the requirement of unitarity directly removes the $a_1$ mode in the free theory, and for the $a_2$ mode it depends on the specific dynamics of the theory. Motivated by this, we remove the normalizable singular mode in the LLL projection.

The $\sigma=0$ flavor has $N=2q+1$ orbitals, while the $\sigma=\pm1$ flavors have $2q$ orbitals. We fix the total fermion number to $N_e=N=2q+1$, and consider the Hamiltonian:
\begin{equation}
     H=\int d^2 r ::n_e(1-V_1\nabla^2)n_e::-U_1 ::\left( n^x n^x + n^y n^y-V_1 n^x \nabla ^2 n^x -V_1n^y \nabla ^2 n^y\right)::+ D \Psi^\dagger (S^z)^2 \Psi.
\end{equation}

The UV symmetry for generic $\nu$ is $U(1)_e\times \left[(SO(2)_r\times U(1)_Q)\rtimes \mathbb Z_{2,RC}\right]$. For generic $\nu$ the twisted boundary condition breaks the $e^{i\pi L_x}$ spatial rotation and the $R_{xz}(\pi)=e^{i\pi S_y}$ spin rotation, the combined transformation $e^{i\pi (L_x+S_y)}$ is preserved and corresponds to the IR $\mathbb Z_{2,RC}$ symmetry, which anti-commutes with both $L_z, Q$. It is sufficient to label the state by $(L_z,Q)$ and restrict to $L_z\geq0$, and the $(0,0)$ sector has an extra $\mathbb Z_2$ quantum number $(0,0)^\pm$.

The phase diagram is discussed in appendix \ref{app: phase transition}. By finite-size scaling of energy levels, we identify a transition at $D_c$, shown in figure \ref{fig: O2_fs_gap}. We use a finite-size cost function $P(U_1,V_1,D,\nu)$ to reduce the irrelevant perturbations, similar to \eqref{eq: fs_cost} , with defect states $|\hat S\rangle,|\hat \Psi_\nu\rangle,|\hat{\bar\Psi}_{1-\nu}\rangle,|\hat D\rangle,|\hat \Psi_{1+\nu}\rangle$. Minimizing $P(U_1,V_1,D,\nu)$ obtained from $N=7,...,11$ at each sampled value of $\nu$, we find critical parameters $U_{1c}(\nu),V_{1c}(\nu),D_c(\nu)$. The conformal cost function $Q(N)$, defined in appendix \ref{app: cost functions}, measures deviations of $\Delta_{PO}-\Delta_O$ from one. As shown in figure \ref{fig:tower_cost}(b), it decrease with $N$ for $N\leq 18$, which provides an independent check of optimization method in that size range. At larger size a residual relevant perturbation from the bulk operator $S$ can have significant effect at large size (see appendix \ref{app: rel_perturbation}), making the determination of the critical point challenging.\footnote{In the following, we use parameters obtained from a two-step tuning procedure that reduces the effect of relevant perturbations at large $N$, see appendix \ref{app: rel_perturbation}.}

\subsection{Spectrum}
\begin{figure}[t]
    \centering
    \begin{minipage}[t]{0.45\linewidth}
    \stackinset{l}{2mm}{t}{2mm}{\textbf{a.}}{
        \includegraphics[width=\linewidth]{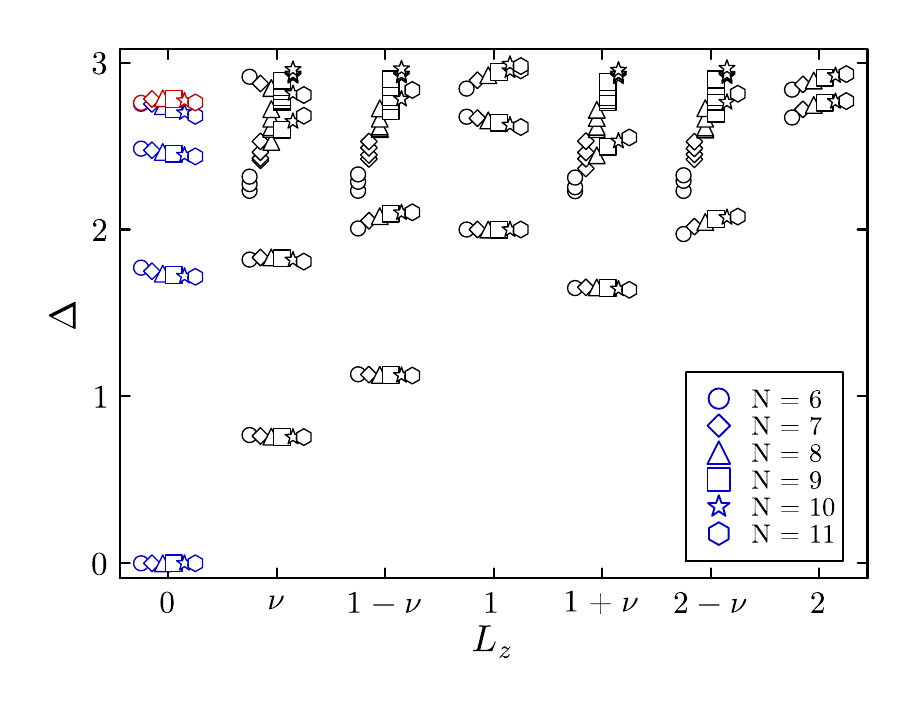}
        }
    \end{minipage}
     \begin{minipage}[t]{0.45\linewidth}
    \stackinset{l}{2mm}{t}{2mm}{\textbf{b.}}{
        \includegraphics[width=\linewidth]{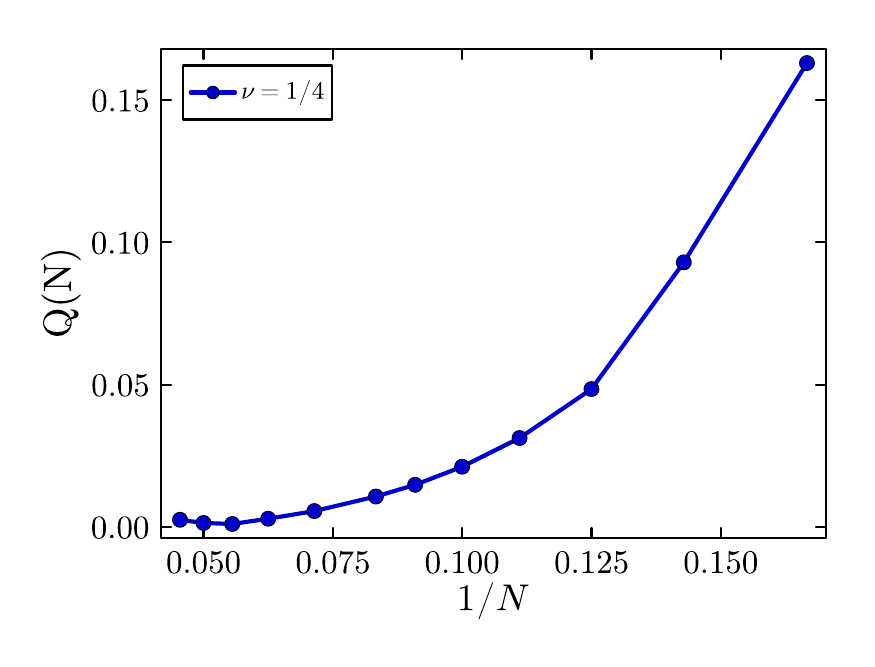}
        }
    \end{minipage}
  
    \caption{(a) Conformal towers at the tuned parameters for $N=6,...,11$. The $\mathbb Z_{2,RC}$ quantum number is defined only in $(0,0)$ sector. Blue points have the same $\mathbb Z_{2,RC}$ quantum number as the vacuum, red points have opposite $\mathbb Z_{2,RC}$ quantum number. (b) Conformal cost function up to $N=22$, both panels show results for $\nu = 1/4$.}
    \label{fig:tower_cost}
\end{figure}

We fix the overall scaling by setting $\Delta_D=2$ and obtain the conformal towers shown in figure \ref{fig:tower_cost}. We then compute the scaling dimensions as a function of $\nu$. At $\nu =1/2$, there is an extra charge conjugation symmetry which exchanges $\Psi_{\nu}$ and $\bar \Psi_{1-\nu}$. At $\nu=0$ the monodromy line defect become transparent, and the spatial $SO(3)_r$ and charge conjugation are restored. In this limit, $\Psi_{\nu}=\phi$, $\bar \Psi_{1-\nu}=\partial_+\phi^*$ and $ \Psi_{1+\nu}=\partial_+\phi$. However there is $1/N$ effect coming from the missing modes $\psi_{1,q}, \psi_{-1,-q}$ at $\nu=0$, that presumably accounts for the observed mismatch in this limit: $\Delta_{\Psi_\nu}$ differs from the bulk $\Delta_\phi$ by 0.05, while the dimensions of $\Delta_{\bar \Psi_{1-\nu}}$ and $ \Delta_{\Psi_{1+\nu}}$ which should become degenerate in the transparent limit, differ by about 0.1.

\begin{figure}[t]
    \centering
    \begin{minipage}[t]{0.45\linewidth}
    \stackinset{l}{2mm}{t}{2mm}{\textbf{a.}}{
        \includegraphics[width=\linewidth]{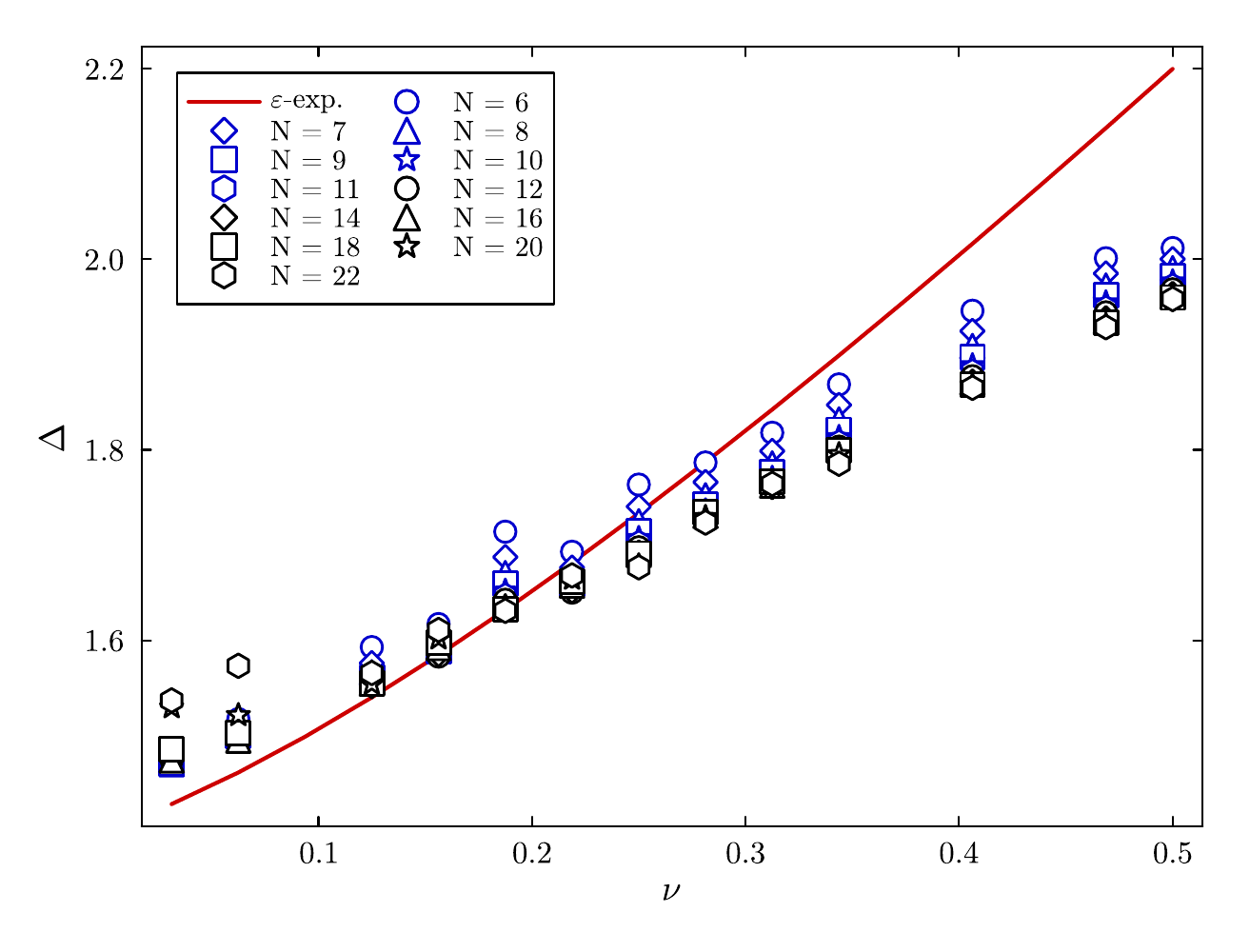}
        }
    \end{minipage}
     \begin{minipage}[t]{0.45\linewidth}
    \stackinset{l}{2mm}{t}{2mm}{\textbf{b.}}{
        \includegraphics[width=\linewidth]{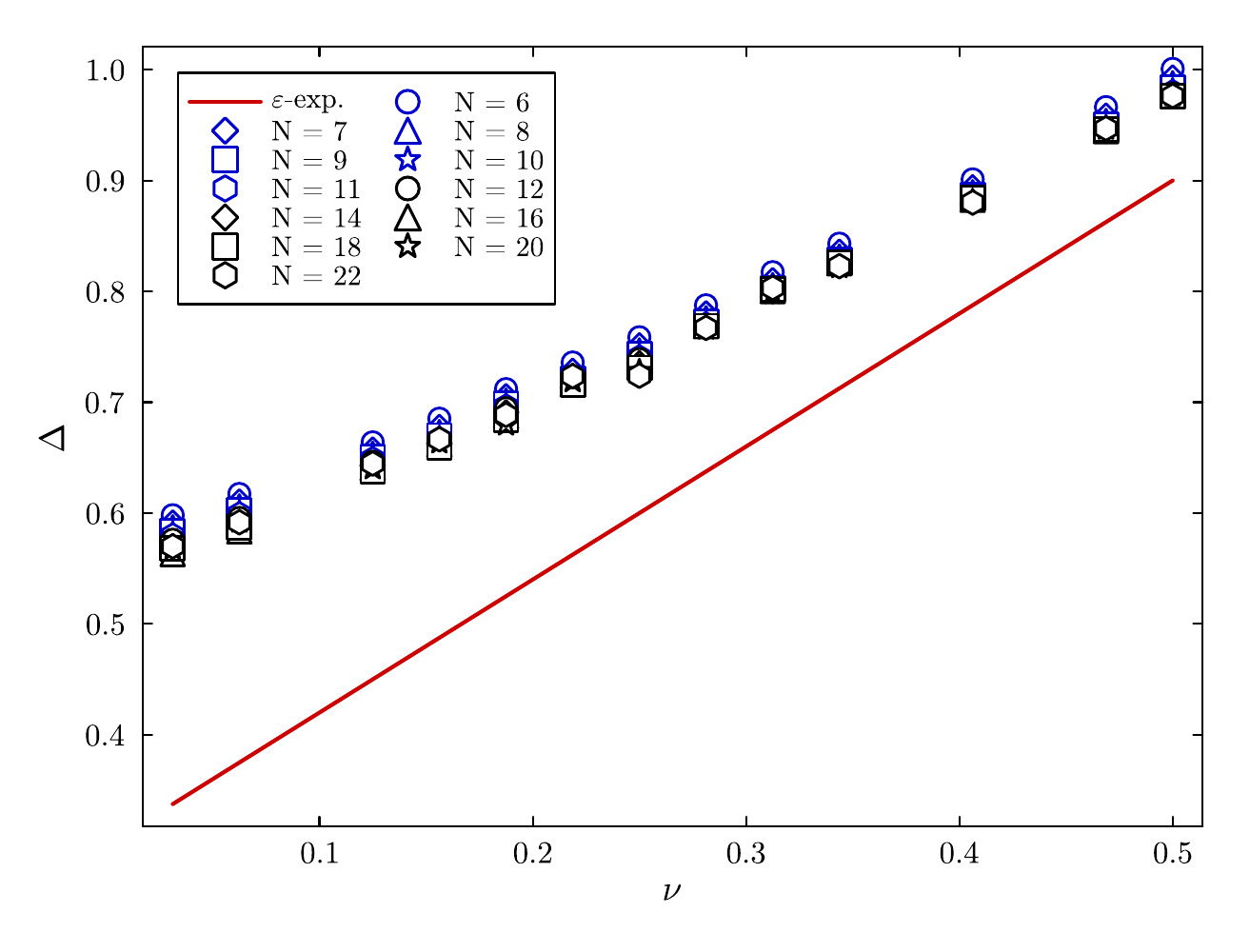}
        }
    \end{minipage}
    \begin{minipage}[t]{0.45\linewidth}
    \stackinset{l}{2mm}{t}{2mm}{\textbf{c.}}{
        \includegraphics[width=\linewidth]{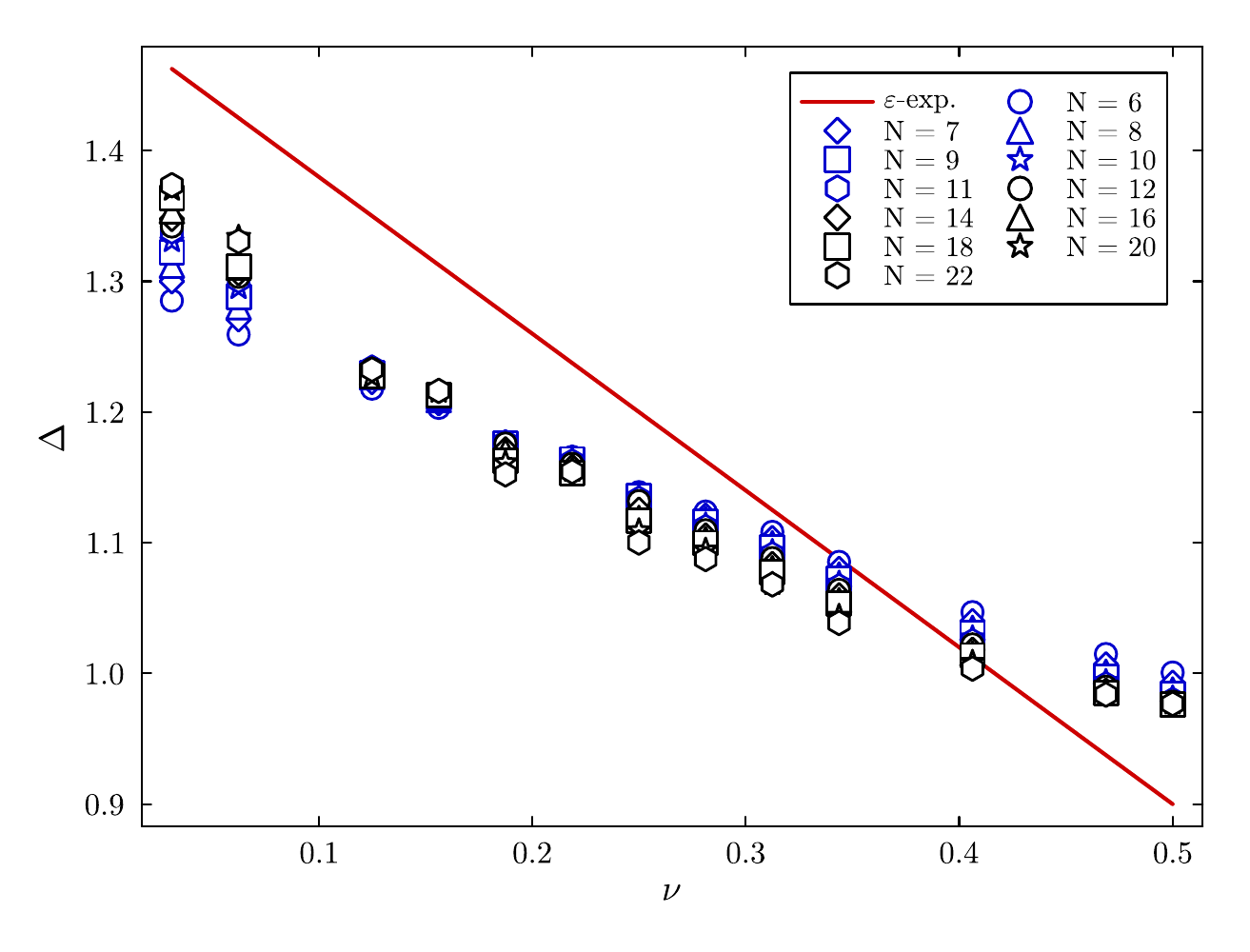}
        }
    \end{minipage}
    \begin{minipage}[t]{0.45\linewidth}
    \stackinset{l}{2mm}{t}{2mm}{\textbf{d.}}{
        \includegraphics[width=\linewidth]{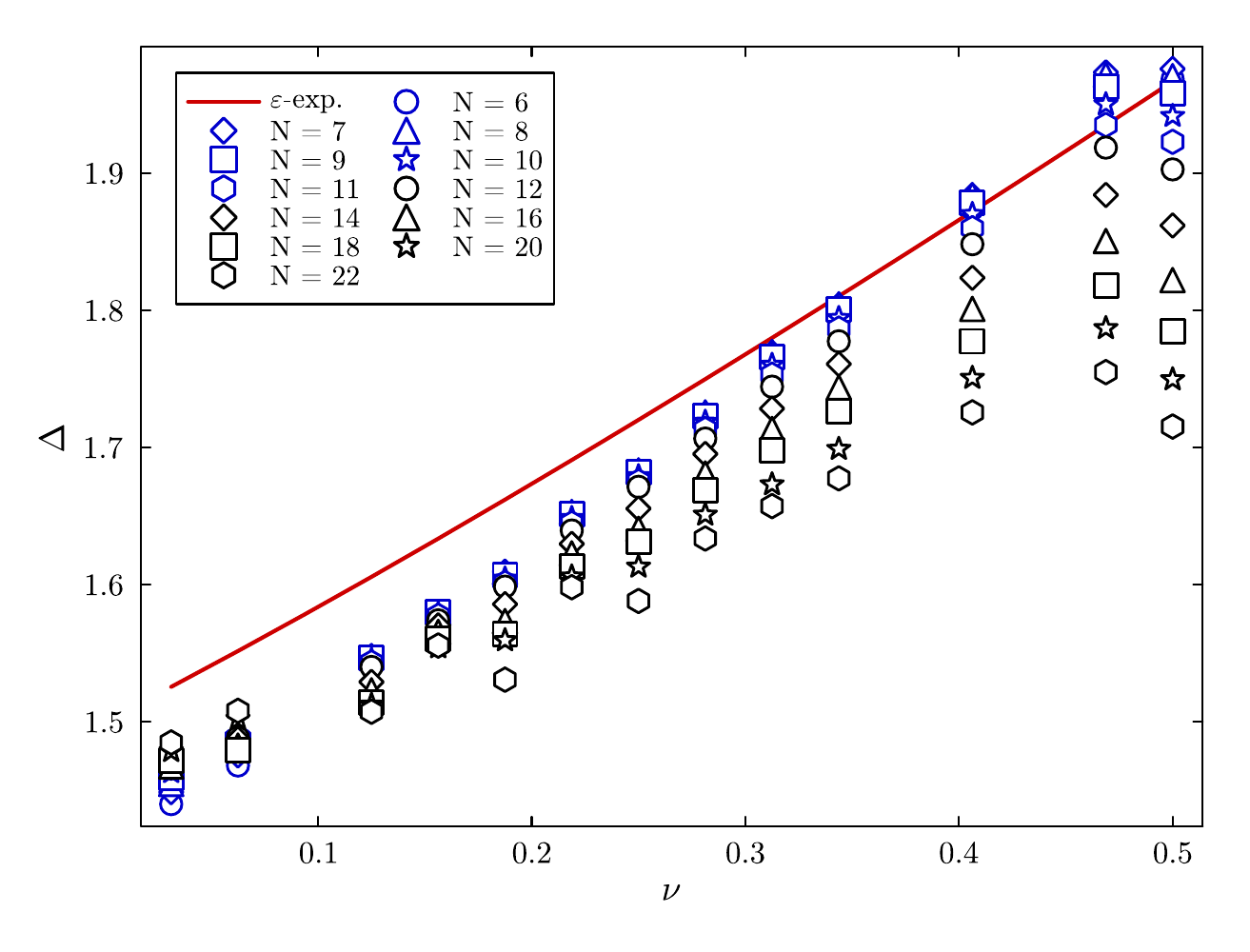}
        }
    \end{minipage}

    \caption{Scaling dimensions as a function of $\nu$ for operator (a) $\hat S$, (b) $\hat \Psi_{\nu}$,  (c) $\hat {\bar\Psi}_{1-\nu}$, and (d) $\hat \Psi_{1+\nu}$. Red curves show the $\varepsilon$-expansion prediction. Results for $N=6,...,11$ are obtained by ED, and those for $N=12,...,22$  by DMRG. }
    \label{fig:O2_spec}
\end{figure}

The results are shown in table \ref{tab:dmrg_N18} at $N=18$, where $Q(N)$ is minimized. Figure \ref{fig:O2_spec} shows their dependence on $\nu$.
For all sampled values of $\nu\in(0,\frac{1}{2}]$, the leading neutral defect scalar $\hat S$ has scaling dimension greater than one. This supports the absence of defect neutral scalar relevant perturbations, which hints that the insertion of a monodromy defect in the UV, for example, using the substitution \eqref{eq: frustrated_link} on a topological surface in the critical $O(2)$ lattice model, will flow to a dCFT without extra defect fine-tuning.

\subsection{DOE coefficients}

We also compute ratios of DOE coefficients. As in \eqref{eq: DOE coefficients}, matching the UV and IR symmetry quantum numbers gives:
\begin{equation}
    \begin{split}
        n^+(\Omega) &= c_{\phi} \phi(\Omega)+...\\
        n^-(\Omega) &= c_{\phi}^* \phi^*(\Omega)+...\\
        n_{z2}(\Omega) &= c_{0}+ c_{S } S(\Omega)+...,
    \end{split}
\end{equation}
where $n^+(\Omega) = \Psi^\dagger S^+ \Psi$, $n^-(\Omega) = \Psi^\dagger S^- \Psi$, $n_{z2}(\Omega) = \Psi^\dagger S_{z}^2 \Psi$. These expansions allow us, in principle, to compute $b_{\phi  \Psi_{\nu-1}}/b_{\phi  \Psi_{\nu-2}}$ and $b_{S \hat S}/b_{S \hat D}$.\footnote{The other ratio $b_{\phi^* \bar\Psi_{-\nu}}/b_{\phi^*\bar  \Psi_{-\nu-1}}$ does not agree with the CFT prediction in our numerical results. This may related to the fact that $\Delta_{\hat\Psi_\nu}$ is close to $\Delta_\phi$, $\langle \hat I| n_-(\theta)|\hat \Psi_{\nu}\rangle$ decays slowly near the defect.} \\

\begin{figure}[t]
    \centering
    \begin{minipage}[t]{0.45\linewidth}
    \stackinset{l}{2mm}{t}{2mm}{\textbf{a.}}{
        \includegraphics[width=\linewidth]{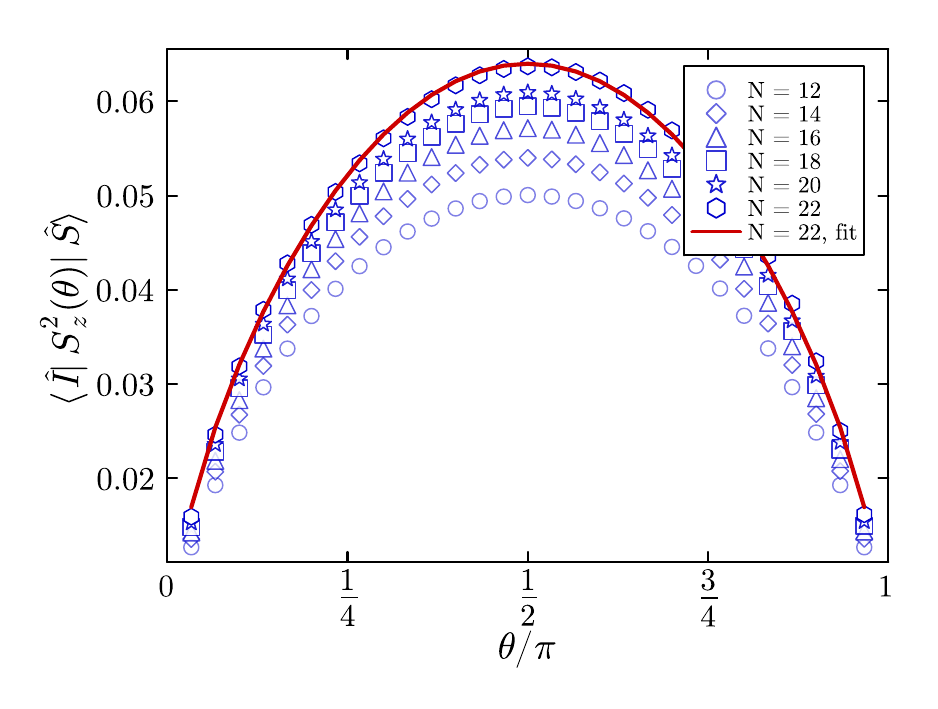}
        }
    \end{minipage}
     \begin{minipage}[t]{0.45\linewidth}
    \stackinset{l}{2mm}{t}{2mm}{\textbf{b.}}{
        \includegraphics[width=\linewidth]{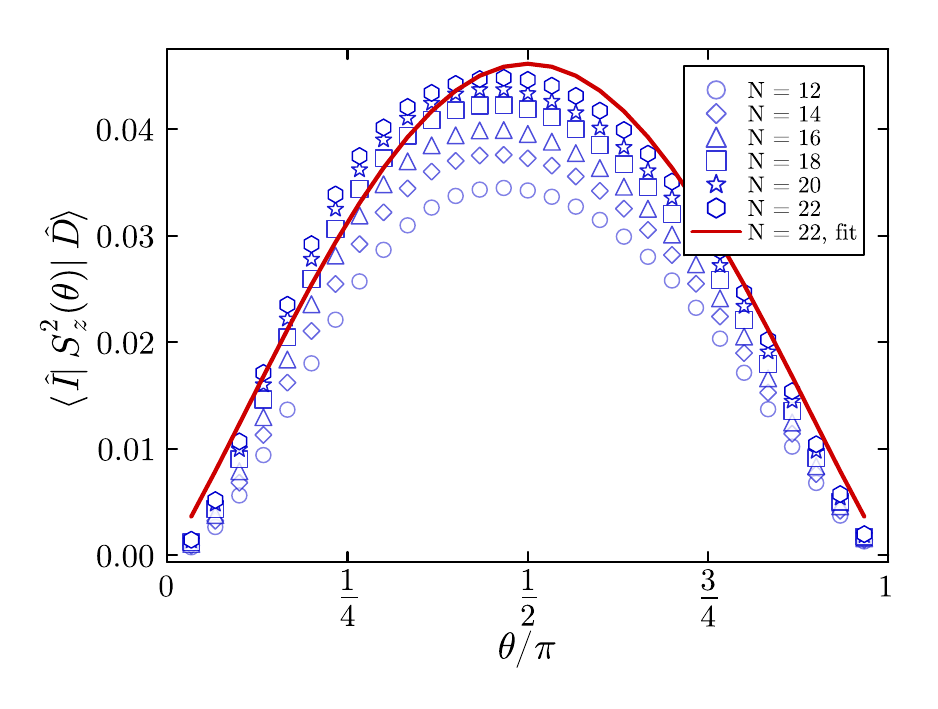}
        }
    \end{minipage}
    \begin{minipage}[t]{0.45\linewidth}
    \stackinset{l}{2mm}{t}{2mm}{\textbf{c.}}{
        \includegraphics[width=\linewidth]{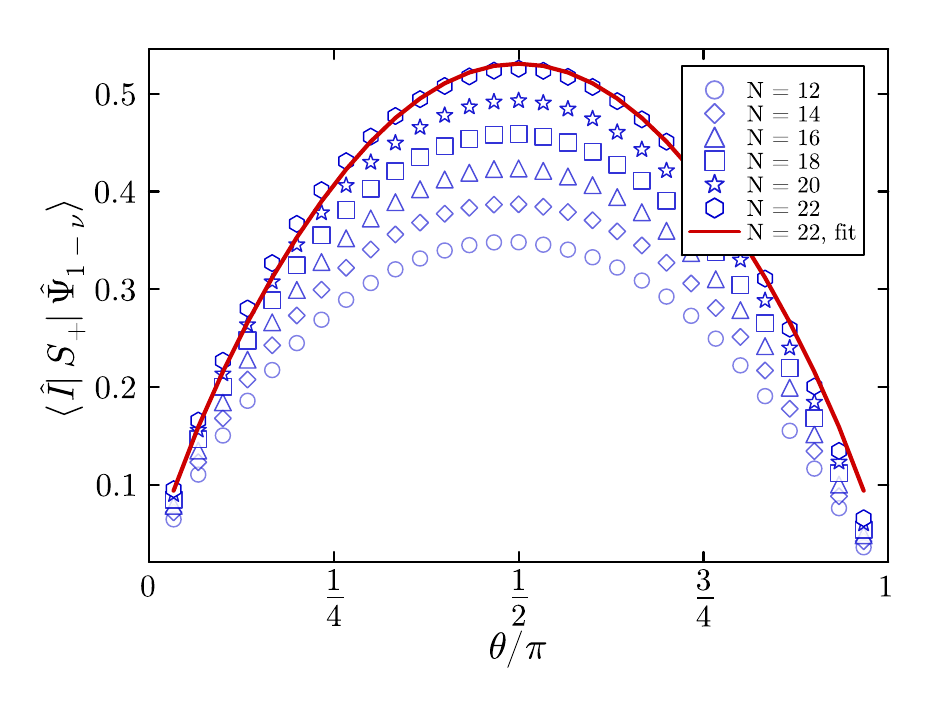}
        }
    \end{minipage}
    \begin{minipage}[t]{0.45\linewidth}
    \stackinset{l}{2mm}{t}{2mm}{\textbf{d.}}{
        \includegraphics[width=\linewidth]{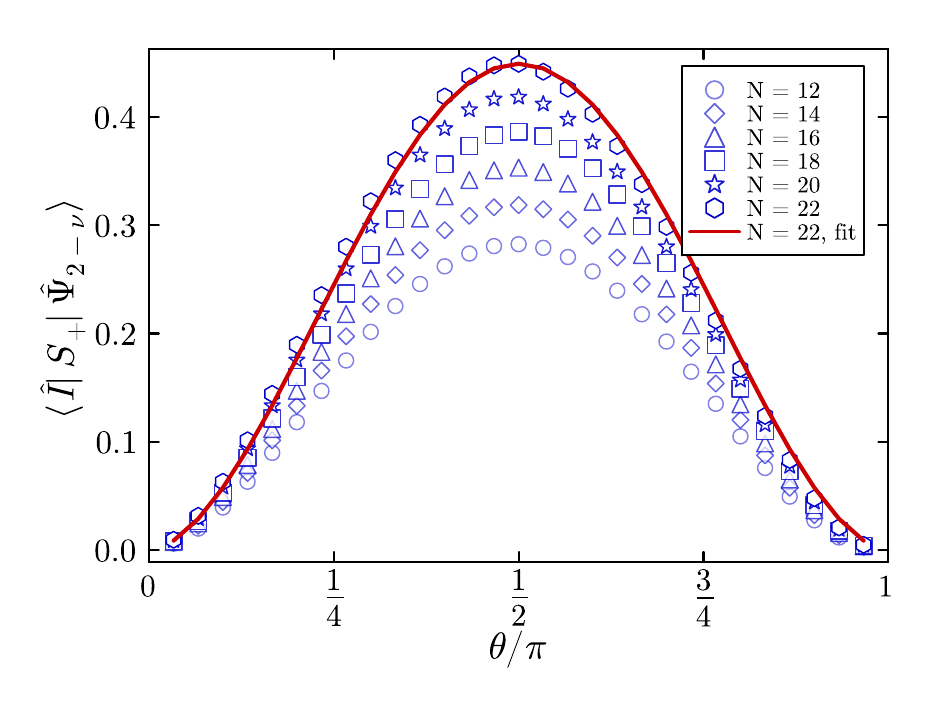}
        }
    \end{minipage}

    \caption{Vacuum to primary matrix elements (a) $\langle \hat I|n_{z2}(\theta,0)|\hat S\rangle$, (b) $\langle \hat I|n_{z2}(\theta,0)|\hat D\rangle$, (c) $\langle \hat I|n_{+}(\theta,0)|\hat {\bar\Psi}_{1-\nu}\rangle$, (d) $\langle \hat I|n_{+}(\theta,0)|\hat {\bar\Psi}_{2-\nu}\rangle$. Red curve shows the CFT prediction. All the results are for $\nu = 1/4$.}
    \label{fig: O2_1pt_mat_ele}
\end{figure}

\begin{figure}[t]
    \centering
    \begin{minipage}[t]{0.45\linewidth}
    \stackinset{l}{2mm}{t}{2mm}{\textbf{a.}}{
        \includegraphics[width=\linewidth]{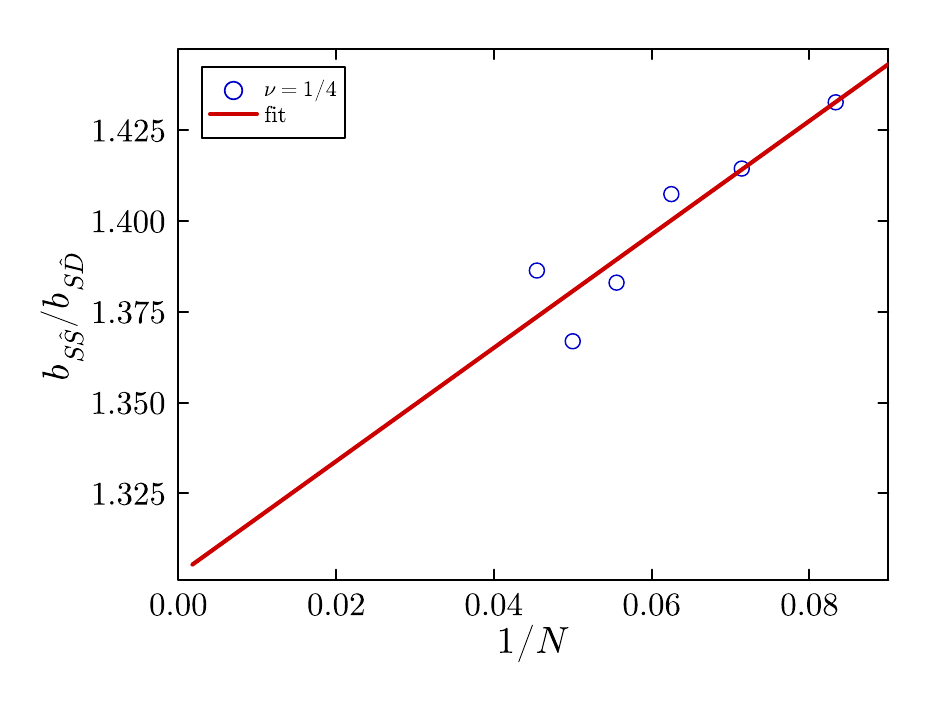}
        }
    \end{minipage}
     \begin{minipage}[t]{0.45\linewidth}
    \stackinset{l}{2mm}{t}{2mm}{\textbf{b.}}{
        \includegraphics[width=\linewidth]{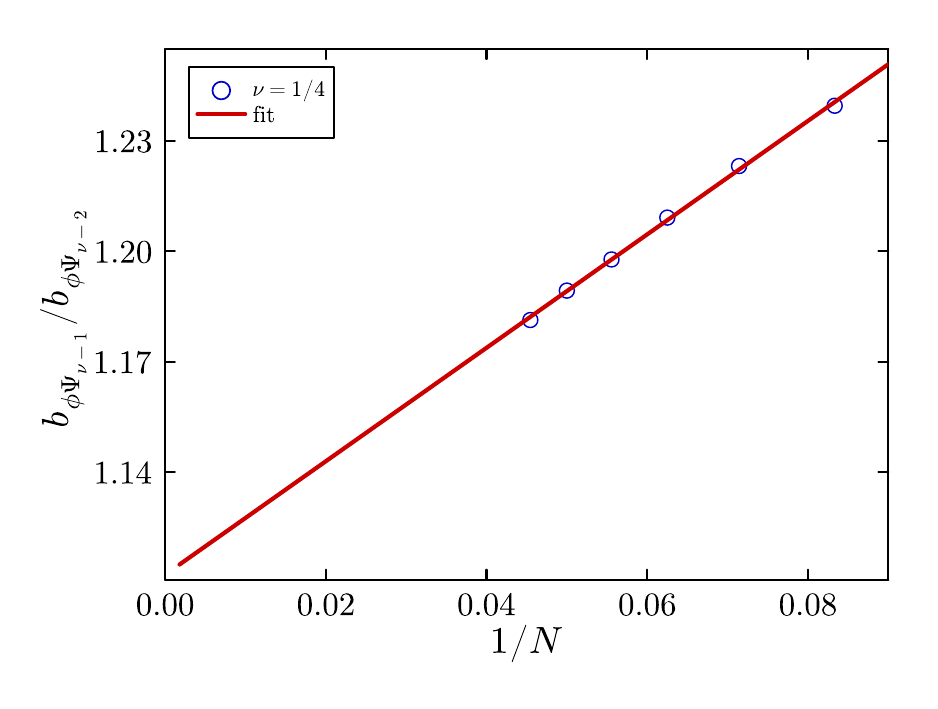}
        }
    \end{minipage}

    \caption{Finite-size extrapolations of DOE-coefficient ratios extracted from vacuum to primary matrix elements: (a) $b_{S \hat S}/b_{S \hat D}$ (b) $b_{\phi  \Psi_{\nu-1}}/b_{\phi  \Psi_{\nu-2}}$. The fits use $b_{S \hat S}/b_{S \hat D}=a+b/N$. All the results are for $\nu = 1/4$.}
    \label{fig: DOE_ratio_fitting}
\end{figure}

Figure \ref{fig: O2_1pt_mat_ele} shows representative matrix elements $\langle \hat I| n(\Omega)|\hat O\rangle $ obtained by DMRG at $N=12,...,22$, together with the corresponding CFT predictions. We extrapolate the ratios $b_{\phi  \Psi_{\nu-1}}/b_{\phi  \Psi_{\nu-2}}$, $b_{S \hat S}/b_{S \hat D}$ using fits of the form $a+b/N$. The leading $1/N$ correction arises from the bulk-descendants contributions $n_+(\Omega) = c_{\phi} \phi(\Omega)+c_{\Box\phi} \Box\phi(\Omega)+...$ of the UV to IR operator expansion. Figure \ref{fig: DOE_ratio_fitting} shows the finite-size extrapolations at $\nu = 1/4$. Applying the same method at other sampled values of $\nu$ gives the ratios listed in table \ref{tab:ratio_fit}.

\section{Discussions}
We construct a fuzzy sphere model for the monodromy line defect in three-dimensional Ising CFT, and extracted defect scaling dimensions and ratios of bulk-to-defect operator expansion coefficients. Our result further supports a defect fixed point with one relevant symmetry-preserving bulk perturbation $\epsilon$, and no relevant symmetry-preserving defect perturbation. We compare our result with MC simulation. We also extend this machinery to the $O(2)$ monodromy line defects. The result support a continuous family of defect CFTs with one relevant symmetry-preserving bulk perturbation, and are consistent with the $\varepsilon-$expansion and bootstrap results.

A direct Monte Carlo study of the three-dimensional \(O(2)\) model with a twisted line defect would provide an independent test of $O(2)$ monodromy line dCFTs. The monodromy could be implemented across a branch surface ending on the line, while correlations along and around the defect could be used to extract defect scaling dimensions. Such measurements would be particularly useful for cross check with the finite-size extrapolation, and will also shed light on the defect bootstrap program.

Under the particle/vortex duality \cite{Peskin:1977kp,Dasgupta:1981zz,Karch:2016sxi,Seiberg:2016gmd} of the XY model and the $U(1)_0+\phi$ scalar QED, an O(2) monodromy defect can be represented by $dB = 2\pi \nu \delta^{2}(x_\perp)$. In the QED side the global O(2) symmetry is the topological symmetry, which couples to the background field by BF coupling $\frac{1}{2\pi}adB$, with the $B$ background, it insert a Wilson line of fractional charge, as studied in non-compact QED \cite{Metlitski:2007fu}. Fractional charged objects are also present in the Chern-Simons/FQH system, to be specific, in $U(1)_0+\phi$ QED (with gauge field $a$), if one adds a BF type coupling of the gauge field $a$ to another gauge field $b$ with CS level $k$ and inserts an integer charge for $b$, from the equation of motion of $b$, one obtain effectively a Wilson line $W_{n/k}(L)$.


Another interesting direction is to realize the monodromy pinning flows on the fuzzy sphere. Perturbation by operator $\Psi_\nu$ or $\bar \Psi_{1-\nu}$ break both the transverse rotation and the internal $U(1)$ rotation. The large-$N$ analysis of ref. \cite{Kravchuk:2025evf} identifies a threshold $\nu^*$ separating the pinning phase (with $\Delta<1$) and the non-pinning phase (with $\Delta>1$). Our results indicate that a relevant deformation $\Psi_\nu$ that can initiate a pinning flow is likely to exist for all $\nu \in[0,\frac{1}{2}]$. However, such a scenario requires further checks, especially near $\nu \rightarrow\frac{1}{2}$.

\acknowledgments{The author thanks Wei Zhu and Shai Chester for useful discussions, and Ofer Aharony for useful discussions and helps on organizing the manuscript. The numerical computations have used the publicly available Julia package FuzzifiED \cite{Zhou_25pkg}, ITensor \cite{Fishman:2020itensor, Schollwock:2010zz}. ChatGPT (OpenAI) was used to assist with language editing and code checking during the preparation of this manuscript. This work was supported in part by ISF grant no.2159/22, by Simons Foundation grant 994296 (Simons Collaboration on Confinement and QCD Strings), by the Minerva foundation with funding from the Federal German Ministry for Education and Research, and by the German Research Foundation through a German-Israeli Project Cooperation (DIP) grant “Holography and the Swampland”. The numerical computations were performed using the WEXAC high-performance computing cluster at Weizmann Institute of Science.\\

\textit{Note added.} During the preparation of this manuscript, we become aware of work \cite{tang2026}, which used the same UV parameter as the bulk model, find the monodromy Ising dCFT and the monodromy $O(2)$ dCFT with twist angle $\pi$, whereas in this work, we find the UV parameters independently by optimize the energy scaling $E_n-E_0 \sim 1/\sqrt{N}$. Also, we are aware of works \cite{MCheng2026a,MCheng2026b} using theoretical and numerical methods to study the monodromy line defect in $O(2)$ WF theory.
}

\appendix

\section{Monopole harmonics with twisted boundary conditions}\label{app: solve_twist_modes}
We take the following ansatz for solutions of the Schr\"odinger equation under a $U(1)$ monopole background,
\begin{equation}
    \begin{split}
Y_{lm}^{(q)}\left(\theta,\phi\right)&=\Theta_{q,l,m}\left(\theta\right)e^{i(m+q)\phi},
    \end{split}
\end{equation}
$\Theta(x)$ satisfies:
\begin{equation}
    \left(l(l+1)-q^{2}\right)\Theta\left(x\right)=-(1-x^{2})\Theta^{''}(x)+2x\Theta'(x)+\frac{1}{1-x^{2}}(m+qx)^{2}\Theta(x).
\end{equation}
The general solution for $\Theta(x)$ which is regular at $x\rightarrow1$ reads:
\begin{equation}
\Theta_{l,m}^{(q)}\left(x\right)\propto\left(1-x\right)^{\frac{q+m}{2}}\left(1+x\right)^{\frac{-q+m}{2}}{}_{2}F_{1}\left(-l+m,1+m+l,q+m+1,\frac{1-x}{2}\right).
\end{equation}
Near the endpoints $x\rightarrow\pm 1$, the asymptotic solutions are:
\begin{equation}
    \begin{split}
        \Theta\left(x\right)&\simeq A_{-}\left(1+x\right)^{-\frac{q-m}{2}}+A_{+}\left(1+x\right)^{\frac{q-m}{2}},\\\Theta\left(x\right)&\simeq B_{-}\left(1-x\right)^{-\frac{m+q}{2}}+B_{+}\left(1-x\right)^{\frac{m+q}{2}}.
    \end{split}
\end{equation}
Regularity requires the coefficients $A_-$ and $B_-$ vanish, using a Hypergeometric connection formula, we obtain:
\begin{equation}
    \begin{split}
\Theta_{l,m}^{(q)}\left(x\right)&\propto\left(1-x\right)^{\frac{q+m}{2}}\left(1+x\right)^{\frac{-q+m}{2}}\frac{\Gamma\left(q+m+1\right)\Gamma\left(q-m\right)}{\Gamma\left(q+1+l\right)\Gamma\left(q-l\right)}\\
        &{}_{2}F_{1}\left(-l+m,1+m+l,2m+2-q-m-1,\frac{1+x}{2}\right)\\&+\left(1-x\right)^{\frac{q+m}{2}}\left(1+x\right)^{\frac{q-m}{2}}\frac{1}{2^{q-m}}\frac{\Gamma\left(q+m+1\right)\Gamma\left(m-q\right)}{\Gamma\left(-l+m\right)\Gamma\left(1+m+l\right)}\\
        &{}_{2}F_{1}\left(q+l+1,q-l,1+q-m,\frac{1+x}{2}\right).
    \end{split}
\end{equation}
Expanding near $x=-1$, regularity requires the first term vanish:
\begin{equation}
     A_{-}=\frac{\Gamma\left(q+m+1\right)\Gamma\left(q-m\right)}{\Gamma\left(q+1+l\right)\Gamma\left(q-l\right)}=0.
\end{equation}
The spectrum reads:
\begin{equation}
    l=q+n, \qquad n\in \mathbb Z_{\geq0},
\end{equation}
and $E_n= \frac{l(l+1)-q^2}{2Mr^2}$, the lowest Landau level projection retains the $n=0$ modes in the $M\rightarrow0$ limit. The mode function reads:
\begin{equation}
\begin{split}
    Y_{q,m}^{(q)}\left(x,\phi\right)&=\mathcal{N}_{q,m}(1-x)^{\frac{m+q}{2}}(x+1)^{\frac{q-m}{2}}e^{i(m+q)\phi},\\
    \mathcal{N}_{q,-m}&=  \mathcal{N}_{q,m} = \sqrt{\frac{2^{-2q-1}}{2\pi} \frac{\Gamma(2q+2)}{\Gamma(q-m+1)\Gamma(q+m+1)}},\label{eq: monopole harmonics}
\end{split}
\end{equation}
with $E_0=\frac{q}{2Mr^2}$. Regularity and monodromy condition under $\phi\rightarrow\phi +2\pi$ require $m=-q+\frac{1}{2},...,q-\frac{1}{2}$.\\
There are also normalizable but singular terms with $m=-q-\frac{1}{2}$, and $m=q+\frac{1}{2}$, which become singular at $x\rightarrow1^-$, $x\rightarrow-1^+$ respectively, for example, for the first mode, $\int_0^\pi d\theta \sin\theta |Y^{(q)}_{q,m}|=\mathcal N_{q,m}^2 2^{2q+\frac{1}{2}}\int_{1-\epsilon}^1dx (1-x)^{-1/2}+\text{reg...}$, the integral converges. However, neither the north-patch nor the south-patch function is single-valued:
\begin{equation}
\begin{split}
    Y^{(N)}(x,\phi)&\sim (1-x)^{-\frac{1}{4}}e^{-i\frac{1}{2}\phi}\\
    Y^{(S)}(x,\phi)&\sim (1-x)^{-\frac{1}{4}}e^{-i(\frac{1}{2}+2q)\phi},
\end{split} 
\end{equation}
we therefore exclude these modes. 

We then consider the $O(2)$ twisted line defects, for the fermion flavor $\psi_{1}$, $m=-q+\nu+\mathbb Z$, the condition:
\begin{equation}
     \frac{\Gamma\left(q+m+1\right)\Gamma\left(q-m\right)}{\Gamma\left(q+1+l\right)\Gamma\left(q-l\right)}=0
\end{equation}
still holds, thus with:
\begin{equation}
    \begin{split}
        l&=q+n, \qquad n\in \mathbb Z_{\geq0}\\
        E_n&= \frac{l(l+1)-q^2}{2Mr^2},
    \end{split}
\end{equation}
one can analytically continue the antiperiodic monopole harmonics \eqref{eq: monopole harmonics} as a function of $m$. There are $2q$ modes, $m=-q+\nu,...,q-1+\nu$, similarly the singular normalizable functions at $m=-q+\nu -1, q+\nu$ are not single-valued in neither patch and are therefore excluded.

We next derive the twisted spherical harmonics used as a complete basis:
\begin{equation}
    \begin{split}
Y_{lm}^{(A)}\left(\theta,\phi\right)&=\Theta_{l,m}\left(\theta\right)e^{im\phi},
    \end{split}
\end{equation}
$\Theta(x)$ satisfy:
\begin{equation}
    l(l+1)\Theta\left(x\right)=-(1-x^{2})\Theta^{''}(x)+2x\Theta'(x)+\frac{1}{1-x^{2}}m^{2}\Theta(x),
\end{equation}
which is a special case of the solution above:
\begin{equation}
\Theta_{l,m}\left(x\right)\propto\left(1-x\right)^{\frac{|m|}{2}}\left(1+x\right)^{\frac{|m|}{2}}{}_{2}F_{1}\left(-l+|m|,1+|m|+l,|m|+1,\frac{1-x}{2}\right).
\end{equation}
Here we have used the invariance of the equation under $m\rightarrow -m$, this solution is regular at $x=1$, while at $x=-1$, using a connection formula for $_2F_1$,
\begin{equation}
    \begin{split}
        \Theta_{l,m}\left(x\right)&\propto\left(1-x\right)^{\frac{|m|}{2}}\left(1+x\right)^{\frac{|m|}{2}}\frac{\Gamma\left(|m|+1\right)\Gamma\left(-|m|\right)}{\Gamma\left(1+l\right)\Gamma\left(-l\right)}{}_{2}F_{1}\left(-l+|m|,1+|m|+l,2|m|+2-|m|-1,\frac{1+x}{2}\right)\\&+\left(1-x\right)^{\frac{|m|}{2}}\left(1+x\right)^{\frac{-|m|}{2}}\frac{2^{|m|}\Gamma\left(|m|+1\right)\Gamma\left(|m|\right)}{\Gamma\left(-l+|m|\right)\Gamma\left(1+|m|+l\right)}{}_{2}F_{1}\left(l+1,-l,1-|m|,\frac{1+x}{2}\right).
    \end{split}
\end{equation}
Regularity at the south pole requires the following coefficient to vanish:
\begin{equation}
    \frac{\Gamma\left(|m|+1\right)\Gamma\left(|m|\right)}{\Gamma\left(-l+|m|\right)\Gamma\left(1+|m|+l\right)}.
\end{equation}
We choose the branch $l>-\frac{1}{2}$, for functions satisfying monodromy condition $Y(\theta,\phi+2\pi) = e^{2\pi i \nu}Y(\theta,\phi) $, there are two cases:
\begin{itemize}
    \item $m=\nu+n$, $n\geq0$, $l=\nu+n+k,k\in\mathbb Z_{\geq0}$...,
    \item $m=\nu+n$, $n<0$, $l=-n-\nu+k,k\in\mathbb Z_{\geq0}$....
\end{itemize}
The solutions are expressed in terms of Jacobi polynomials for the case $l-|m|\in\mathbf N$,
\begin{equation}
    Y_{lm}^{(\nu)}\left(\theta,\phi\right)=2^{-|m|}\sqrt{\frac{l+\frac{1}{2}}{2\pi}\frac{\Gamma(l-|m|+1)\Gamma(l+|m|+1)}{\Gamma(l+1)^{2}}}\left(1-x^{2}\right)^{\frac{|m|}{2}}P_{l-|m|}^{|m|,|m|}\left(x\right)e^{im\phi}.\label{eq: spherical_harmonics_T}
\end{equation}
Here $\nu$ specify the twist of spherical harmonics. For negative twist angle, we obtain:
\begin{equation}
    Y_{lm}^{(-\nu)}\left(\theta,\phi\right)=2^{-|m|}\sqrt{\frac{l+\frac{1}{2}}{2\pi}\frac{\Gamma(l-|m|+1)\Gamma(l+|m|+1)}{\Gamma(l+1)^{2}}}\left(1-x^{2}\right)^{\frac{|m|}{2}}P_{l-|m|}^{|m|,|m|}\left(x\right)e^{im\phi},\label{eq: spherical_harmonics_T_bar}
\end{equation}
\begin{itemize}
    \item $m=-\nu+n$, $n>0$, $l=-\nu+n+k,k\in\mathbb Z_{\geq0}$
    \item $m=-\nu+n$, $n\leq0$, $l=\nu-n+k,k\in\mathbb Z_{\geq0}$.
\end{itemize}

\section{Useful integrals}\label{app: 3Y_integral}
It is useful to compute the spherical integrals of two monopole harmonics and one twisted spherical harmonics with a compatible periodicity, we split in three cases:
\subsection{\texorpdfstring{$P\times P\rightarrow P$}{PP->P}}
We consider all three functions are periodic:
\begin{equation}
    \begin{split}
       C^{q,l,(PPP)}_{m_1,m_2}= \int^{2\pi}_0 d\phi \int_0^\pi d\theta\sin \theta  \bar Y_{q,m_1}^{(q)}(\Omega)  Y_{q,m_2}^{(q)}(\Omega)Y_{l,m}^{(P)}(\Omega),
    \end{split}
\end{equation}
\begin{equation}
    \begin{split}
       C^{q,l,(PPP)}_{m_1,m_2}=  2\pi \delta_{m+m_2-m_1} \mathcal N_{l,m_1-m_2}^{(P)}\mathcal N_{q,m_1}\mathcal N_{q,m_2}
       \int_{-1}^1 dx P_{l,m_1-m_2}(x)(1-x)^{\frac{2q+m_1+m_2}{2}}(1+x)^{\frac{2q-m_1-m_2}{2}},
    \end{split}
\end{equation}
where $m_1,m_2=-q, -q+1,..., q$. We will evaluate this integral below.

\subsection{\texorpdfstring{$T\times  T \rightarrow P$}{TT->P}}
\begin{equation}
    \begin{split}
        C_{m_{1},m_{4}}^{q,l(T TP)}&=\int d^{2}\Omega Y_{q,m_{1}}^{(q,\nu)*}\left(\Omega\right)Y_{q,m_{4}}^{(q,\nu)}\left(\Omega\right)Y_{l,m}^{(P)}\left(\Omega\right),\\
        C_{m_{1},m_{4}}^{q,l(\bar T \bar TP)}&=\int d^{2}\Omega Y_{q,m_{1}}^{(q,-\nu)*}\left(\Omega\right)Y_{q,m_{4}}^{(q,-\nu)}\left(\Omega\right)Y_{l,m}^{(P)}\left(\Omega\right),
    \end{split}
\end{equation}

\begin{equation}
    \begin{split}
C_{m_{1},m_{2}}^{q,l,TTP}=&\Gamma(2q+2)\sqrt{\frac{1}{2\pi}\frac{\left(l+\frac{1}{2}\right)\Gamma\left(l-m_{1}+m_{2}+1\right)}{\Gamma\left(l+m_{1}-m_{2}+1\right)}}\sqrt{\frac{\Gamma(q-m_{2}+1)\Gamma(q+m_{2}+1)}{\Gamma(q-m_{1}+1)\Gamma(q+m_{1}+1)}}\\
        &{}_{3}\tilde{F}_{2}\left(-l,l+1,q+m_{2}+1;1-m_{1}+m_{2},2q+2;1\right),
    \end{split}
\end{equation}
where $m_1,m_2=-q+\nu,...,q-1+\nu $, while for $C_{m_{1},m_{2}}^{q,l(\bar T\bar TP)}$, we take $m_1,m_2=-q+1-\nu,...,q-\nu $. Taking $\nu=0$ we obtain the $C^{q,l(PPP)}_{m_1, m_2}$ case. 

\subsection{\texorpdfstring{$T\times P\rightarrow T, \bar T\times P\rightarrow \bar T$}{TP->T}}
\begin{equation}
    \begin{split}
        C_{m_{1},m_{4}}^{q,l(TPT)}&=\int d^{2}\Omega Y_{q,m_{1}}^{(q,\nu)*}\left(\Omega\right)Y_{q,m_{4}}^{(q,0)}\left(\Omega\right)Y_{l,m}^{(\nu)}\left(\Omega\right),\\m_{1}-m_{4}&>0\qquad l=\nu+n,\\
m_{1}-m_{4}&<0\qquad l=1-\nu+n.\\
    \end{split}
\end{equation}
We denote the function with positive twist angle i.e. $Y(\theta,\phi+2\pi) = e^{2\pi i \nu}Y(\theta,\phi) $, $\nu\in(0,\frac{1}{2}]$ by $T$ and those with negative twist by $\bar T$, the barred integral is,
\begin{equation}
    \begin{split}
        C_{m_{1},m_{4}}^{q,l(\bar TP\bar T)}&=\int d^{2}\Omega Y_{q,m_{1}}^{(q,-\nu)*}\left(\Omega\right)Y_{q,m_{4}}^{(q,0)}\left(\Omega\right)Y_{l,m}^{(-\nu)}\left(\Omega\right),\\m_{1}-m_{4}&>0\qquad l=1-\nu+n,\\
m_{1}-m_{4}&<0\qquad l=\nu+n.\\
    \end{split}
\end{equation}
One can represent $C^{(P\bar TT)}$ in terms of $C^{(\bar TP\bar T)}$ with:
\begin{equation}
    Y^{(\nu)*}_{l,m}(\theta,\phi)= Y^{(-\nu)}_{l,-m}(\theta,\phi),
\end{equation}
which gives:
\begin{equation}
    \begin{split}
        &\left[\int d^{2}\Omega Y_{q,m_{4}}^{(q,-\nu)}\left(\Omega\right)Y_{q,m_{1}}^{(q,0)*}\left(\Omega\right)Y_{l,m_{1}-m_{4}}^{(\nu)}\left(\Omega\right)\right]^{*}\\=&\int d^{2}\Omega Y_{q,m_{4}}^{(q,-\nu)*}\left(\Omega\right)Y_{q,m_{1}}^{(q,0)}\left(\Omega\right)Y_{l,m_{1}-m_{4}}^{(\nu)*}\left(\Omega\right)\\=&\int d^{2}\Omega Y_{q,m_{4}}^{(q,-\nu)*}\left(\Omega\right)Y_{q,m_{1}}^{(q,0)}\left(\Omega\right)Y_{l,m_{4}-m_{1}}^{(-\nu)}\left(\Omega\right)\\=&C_{m_{4},m_{1}}^{q,l(\bar{T}P\bar{T})},
    \end{split}
\end{equation}
All the integrals reduce to real integral of $\Theta(x)$, so the $C$ above are real.
Defining $d=|m_{1}-m_{2}|$, we obtain the analytic expression:
\begin{equation}
\begin{split}    C_{m_{1},m_{2}}^{q,l,(TPT)}&=\Gamma(2q+2)\sqrt{\frac{\Gamma(q+\max(m_{1},m_{2})+1)\Gamma(q-\min(m_{1},m_{2})+1)}{\Gamma(q-\max(m_{1},m_{2})+1)\Gamma(q+\min(m_{1},m_{2})+1)}}\sqrt{\frac{\Gamma(l+d+1)}{\Gamma(l-d+1)}}\sqrt{\frac{l+\frac{1}{2}}{2\pi}}\\
&{}_{3}\tilde{F}_{2}\left(d-l,l+d+1,q+\max(m_{1},m_{2})+1;d+1,2q+d+2;1\right).
\end{split}
\end{equation}
$C_{m_{1},m_{2}}^{q,l,(\bar TP\bar T)}$ has the same form with $m_1, m_2, l$ chosen to satisfy the twisted boundary conditions and eigenvalue of the Laplacian equation. For the Ising twist, $\nu=1/2$, the $T$ and $\bar T$ are the same and denoted as $A$.
\section{Fuzzy-sphere Hamiltonian}
\subsection{Ising model}\label{app: Ising Hamiltonian}
We start from a local Hamiltonian of a form:
\begin{equation}
    \begin{split}
        H_{\text{int}}=&\frac{R^{2}}{2}\int d^{2}\Omega ::n^{0}\left(\vec{r}\right)\left(V_{0}+\frac{V_{1}}{R^{2}}\left(-\nabla^{2}\right)+...+\frac{V_{k}}{R^{2k}}\left(-\nabla^{2}\right)^{k}\right)n^{0}\left(\vec{r}\right)::\\-&\frac{R^{2}}{2}\int d^{2}\Omega ::n^{x}\left(\vec{r}\right)\left(V_{0}+\frac{V_{1}}{R^{2}}\left(-\nabla^{2}\right)+...+\frac{V_{k}}{R^{2k}}\left(-\nabla^{2}\right)^{k}\right)n^{x}\left(\vec{r}\right)::.
    \end{split}
\end{equation}
Here in the basis of $\Psi=(\psi_\rightarrow,\psi_\leftarrow)$, where the twisted boundary condition is diagonal. Since $n^x$ is antiperiodic, we use the $T=\bar T =A$ wave function.
\begin{equation}
    \begin{split}
        H_{\text{int}}=&\frac{R^{2}}{2}\sum_{l,m}\left(V_{0}+\frac{V_{1}}{R^{2}}\lambda_{l}+...+\frac{V_{k}}{R^{2k}}\lambda_{l}^{k}\right)\int d^{2}\Omega n^{0}\left(\Omega\right)Y_{lm}^{(P)}\left(\Omega\right)\int d^{2}\Omega'\bar{Y}_{lm}^{(P)}\left(\Omega'\right)n^{0}\left(\Omega'\right)\\-&\frac{R^{2}}{2}\sum_{l,m}\left(V_{0}+\frac{V_{1}}{R^{2}}\lambda_{l}+...+\frac{V_{k}}{R^{2k}}\lambda_{l}^{k}\right)\int d^{2}\Omega n^{x}\left(\Omega\right)Y_{lm}^{(A)}\left(\Omega\right)\int d^{2}\Omega'\bar{Y}_{lm}^{(A)}\left(\Omega'\right)n^{x}\left(\Omega'\right).
    \end{split}
\end{equation}
Using:
\begin{equation}
    \begin{split}
  \bar{Y}_{lm}^{(P)}\left(\Omega\right)&=Y_{l-m}^{(P)}\left(\Omega\right)(-1)^m,\\\bar{Y}_{lm}^{(A)}\left(\Omega\right)&=Y_{l-m}^{(A)}\left(\Omega\right),
    \end{split}
\end{equation}
after doing the mode expansion and performing the 3Y integrals separately for $\Omega$ and $\Omega'$ integrals, we obtain the Hamiltonian in the LLL basis:
\begin{equation}
    \begin{split}
        H_{\text{int}}=&\frac{1}{2N}\sum_{l=0}^{\infty}\sum_{m=-l}^{l}\left(V_{0}+\frac{V_{1}}{N}l\left(l+1\right)+...+\frac{V_{k}}{N^{k}}l^{k}\left(l+1\right)^{k}\right)\\&\sum_{m_{1},m_{2},m_{3},m_{4}=-q}^{q}\psi_{\rightarrow,m_{1}}^{\dagger}\psi_{\rightarrow,m_{2}}^{\dagger}\psi_{\rightarrow,m_{3}}\psi_{\rightarrow,m_{4}}C_{m_{1},m_{4}}^{q,l,PPP}C_{m_{2},m_{3}}^{q,l,PPP}\delta_{\left(m_{4}-m_{1}\right)+\left(m_{3}-m_{2}\right)}\left(-1\right)^{m_4-m_1}\\+&2\times\sum_{m_{1},m_{4}=-q}^{q}\sum_{m_{2},m_{3}=-q+\frac{1}{2}}^{q-\frac{1}{2}}\psi_{\rightarrow,m_{1}}^{\dagger}\psi_{\leftarrow,m_{2}}^{\dagger}\psi_{\leftarrow,m_{3}}\psi_{\rightarrow,m_{4}}C_{m_{1},m_{4}}^{q,l,PPP}C_{m_{2},m_{3}}^{q,l,AAP}\delta_{\left(m_{4}-m_{1}\right)+\left(m_{3}-m_{2}\right)}\left(-1\right)^{m_4-m_1}\\&+\sum_{m_{1},m_{2},m_{3},m_{4}=-q+\frac{1}{2}}^{q-\frac{1}{2}}\psi_{\leftarrow,m_{1}}^{\dagger}\psi_{\leftarrow,m_{2}}^{\dagger}\psi_{\leftarrow,m_{3}}\psi_{\leftarrow,m_{4}}C_{m_{1},m_{4}}^{q,l,AAP}C_{m_{2},m_{3}}^{q,l,AAP}\delta_{\left(m_{4}-m_{1}\right)+\left(m_{3}-m_{2}\right)}\left(-1\right)^{m_4-m_1}\\-&\frac{1}{2N}\sum_{l=\frac{1}{2}}^{\infty}\sum_{m=-l}^{l}\left(V_{0}+\frac{V_{1}}{N}l\left(l+1\right)+...+\frac{V_{k}}{N^{k}}l^{k}\left(l+1\right)^{k}\right)\\&\sum_{m_{1},m_{2}=-q}^{q}\sum_{m_{3},m_{4}=-q+\frac{1}{2}}^{q-\frac{1}{2}}C_{m_{4},m_{1}}^{q,l,APA}C_{m_{3},m_{2}}^{q,l,APA}\psi_{\rightarrow,m_{1}}^{\dagger}\psi_{\rightarrow,m_{2}}^{\dagger}\psi_{\leftarrow,m_{3}}\psi_{\leftarrow,m_{4}}\delta_{\left(m_{4}-m_{1}\right)+\left(m_{3}-m_{2}\right)}\\+&2\times\sum_{m_{1},m_{3}=-q}^{q}\sum_{m_{2},m_{4}=-q+\frac{1}{2}}^{q-\frac{1}{2}}\psi_{\rightarrow,m_{1}}^{\dagger}\psi_{\leftarrow,m_{2}}^{\dagger}\psi_{\rightarrow,m_{3}}\psi_{\leftarrow,m_{4}}C_{m_{4},m_{1}}^{q,l,APA}C_{m_{2},m_{3}}^{q,l,APA}\delta_{\left(m_{4}-m_{1}\right)+\left(m_{3}-m_{2}\right)}\\+&\sum_{m_{1},m_{2}=-q+\frac{1}{2}}^{q-\frac{1}{2}}\sum_{m_{3},m_{4}=-q}^{q}\psi_{\leftarrow,m_{1}}^{\dagger}\psi_{\leftarrow,m_{2}}^{\dagger}\psi_{\rightarrow,m_{3}}\psi_{\rightarrow,m_{4}}C_{m_{1},m_{4}}^{q,l,APA}C_{m_{2},m_{3}}^{q,l,APA }\delta_{\left(m_{4}-m_{1}\right)+\left(m_{3}-m_{2}\right)},\\
        H_x  =h_x&\sum_{m=-q+\frac{1}{2}}^{q-\frac{1}{2}} \psi_{m,\leftarrow}^\dagger \psi_{m,\leftarrow} - h_x\sum_{m=-q}^q \psi_{m,\rightarrow}^\dagger \psi_{m,\rightarrow}.
        \end{split}\label{eq: Hamiltonian_Ising_expansion}
\end{equation}
In this work we retain only $V_0, V_1$ and set $V_0=1$.

\subsection{O(2) model}\label{app: O2_hamiltonian}
The local Hamiltonian is:
\begin{equation}
\begin{split}
     H_{ee}&=\int d^2 r n_e^2 + V_1n_e(-\nabla^2)n_e,\\
     H_{xy}&=\int d^2 rU_1 \left( n^x n^x + n^y n^y-V_1 n^x \nabla ^2 n^x -V_1n^y \nabla ^2 n^y\right),\\
    H_{D}&= \int d^2 rD \Psi^\dagger (S^z)^2 \Psi,\\
    H &= H_{ee}-H_{xy}+H_D.
\end{split}
\end{equation}
The mode expansions are given in \eqref{eq: mode_exp_O2}.
\begin{equation}
    n^a(\Omega) =\frac{1}{R^2}\sum_{m_1,m_2}\sum_{\sigma,\sigma'} \psi^\dagger_{m_1,\sigma}\psi_{m_2,\sigma'} Y_{q,m_2}^{(q)}(\Omega)\bar Y_{q,m_1}^{(q)}(\Omega)S^a_{\sigma\sigma'},
\end{equation}
with $S^a$ the spin-1 $\mathfrak{so}(3)$ generators. After performing the spherical integration using the results of appendix \ref{app: 3Y_integral}, we retain only $V_0,V_1\neq0$ and set $V_0=1$, and then $H_{xy}$ reads:
\begin{equation}
    \begin{split}
        H_{xy}=\frac{2U_1}{N}\sum_{m_{1},m_{2},m_{3},m_{4}}\sum_{l}\left(1+\frac{V_{1}}{N}\lambda_{l}\right)&\psi_{m_{1},1}^{\dagger}\psi_{m_{2},-1}^{\dagger}\psi_{m_{3},0}\psi_{m_{4},0}C_{m_{1},m_{4}}^{q,l(TPT)}C_{m_{2},m_{3}}^{q,l(\bar{T}P\bar{T})}\\+\frac{2U_1}{N}\sum_{m_{1},m_{2},m_{3},m_{4}}\sum_{l}\left(1+\frac{V_{1}}{N}\lambda_{l}\right)&\psi_{m_{1},1}^{\dagger}\psi_{m_{2},0}^{\dagger}\psi_{m_{3},1}\psi_{m_{4},0}C_{m_{1},m_{4}}^{q,l(TPT)}\left[C_{m_{3},m_{2}}^{q,l(TPT)}\right]^{*}\\+\frac{2U_1}{N}\sum_{m_{1},m_{2},m_{3},m_{4}}\sum_{l}\left(1+\frac{V_{1}}{N}\lambda_{l}\right)&\psi_{m_{1},0}^{\dagger}\psi_{m_{2},-1}^{\dagger}\psi_{m_{3},0}\psi_{m_{4},-1}\left[C_{m_{4},m_{1}}^{q,l(\bar{T}P\bar{T})}\right]^{*}C_{m_{2},m_{3}}^{q,l(\bar{T}P\bar{T})}\\+\frac{2U_1}{N}\sum_{m_{1},m_{2},m_{3},m_{4}}\sum_{l}\left(1+\frac{V_{1}}{N}\lambda_{l}\right)&\psi_{m_{1},0}^{\dagger}\psi_{m_{2},0}^{\dagger}\psi_{m_{3},1}\psi_{m_{4},-1}\left[C_{m_{4},m_{1}}^{q,l(\bar{T}P\bar{T})}\right]^{*}\left[C_{m_{3},m_{2}}^{q,l(TPT)}\right]^{*},
    \end{split}
\end{equation}
where $\lambda_l=l(l+1)$, and $l$ corresponds to the values in \eqref{eq: spherical_harmonics_T} and \eqref{eq: spherical_harmonics_T_bar}. The density-interaction term $H_{ee}$ is:
\begin{equation}
    \begin{split}
        H_{ee}=&\frac{1}{N}\sum_{m_{1},m_{2},m_{3},m_{4}}\sum_{l=0}^{l_{\Lambda}}\left(-1\right)^{m_{1}-m_{4}}\left(1+\frac{V_{1}}{N}\lambda_{l}\right)\psi_{m_{1},1}^{\dagger}\psi_{m_{2},1}^{\dagger}\psi_{m_{3},1}\psi_{m_{4},1}C_{m_{1},m_{4}}^{q,l(TTP)}C_{m_{2},m_{3}}^{q,l(TTP)}\\+&\frac{1}{N}\sum_{m_{1},m_{2},m_{3},m_{4}}\sum_{l=0}^{l_{\Lambda}}\left(-1\right)^{m_{1}-m_{4}}\left(1+\frac{V_{1}}{N}\lambda_{l}\right)\psi_{m_{1},0}^{\dagger}\psi_{m_{2},0}^{\dagger}\psi_{m_{3},0}\psi_{m_{4},0}C_{m_{1},m_{4}}^{q,l(PPP)}C_{m_{2},m_{3}}^{q,l(PPP)}\\+&\frac{1}{N}\sum_{m_{1},m_{2},m_{3},m_{4}}\sum_{l=0}^{l_{\Lambda}}\left(-1\right)^{m_{1}-m_{4}}\left(1+\frac{V_{1}}{N}\lambda_{l}\right)\psi_{m_{1},-1}^{\dagger}\psi_{m_{2},-1}^{\dagger}\psi_{m_{3},-1}\psi_{m_{4},-1}C_{m_{1},m_{4}}^{q,l(\bar{T}\bar{T}P)}C_{m_{2},m_{3}}^{q,l(\bar{T}\bar{T}P)}\\+&\frac{2}{N}\sum_{m_{1},m_{2},m_{3},m_{4}}\sum_{l=0}^{l_{\Lambda}}\left(-1\right)^{m_{1}-m_{4}}\left(1+\frac{V_{1}}{N}\lambda_{l}\right)\psi_{m_{1},1}^{\dagger}\psi_{m_{2},0}^{\dagger}\psi_{m_{3},0}\psi_{m_{4},1}C_{m_{1},m_{4}}^{q,l(TTP)}C_{m_{2},m_{3}}^{q,l(PPP)}\\+&\frac{2}{N}\sum_{m_{1},m_{2},m_{3},m_{4}}\sum_{l=0}^{l_{\Lambda}}\left(-1\right)^{m_{1}-m_{4}}\left(1+\frac{V_{1}}{N}\lambda_{l}\right)\psi_{m_{1},-1}^{\dagger}\psi_{m_{2},0}^{\dagger}\psi_{m_{3},0}\psi_{m_{4},-1}C_{m_{1},m_{4}}^{q,l(\bar{T}\bar{T}P)}C_{m_{2},m_{3}}^{q,l(PPP)}\\+&\frac{2}{N}\sum_{m_{1},m_{2},m_{3},m_{4}}\sum_{l=0}^{l_{\Lambda}}\left(-1\right)^{m_{1}-m_{4}}\left(1+\frac{V_{1}}{N}\lambda_{l}\right)\psi_{m_{1},1}^{\dagger}\psi_{m_{2},-1}^{\dagger}\psi_{m_{3},-1}\psi_{m_{4},1}C_{m_{1},m_{4}}^{q,l(TTP)}C_{m_{2},m_{3}}^{q,l(\bar{T}\bar{T}P)}.
    \end{split}
\end{equation}
$l_\Lambda$ is the total angular momentum cut-off, which is taken to be $l_\Lambda=100$.

\section{Phases and phase transitions}\label{app: phase transition}
\begin{figure}[t]
    \centering
    \includegraphics[width=\linewidth]{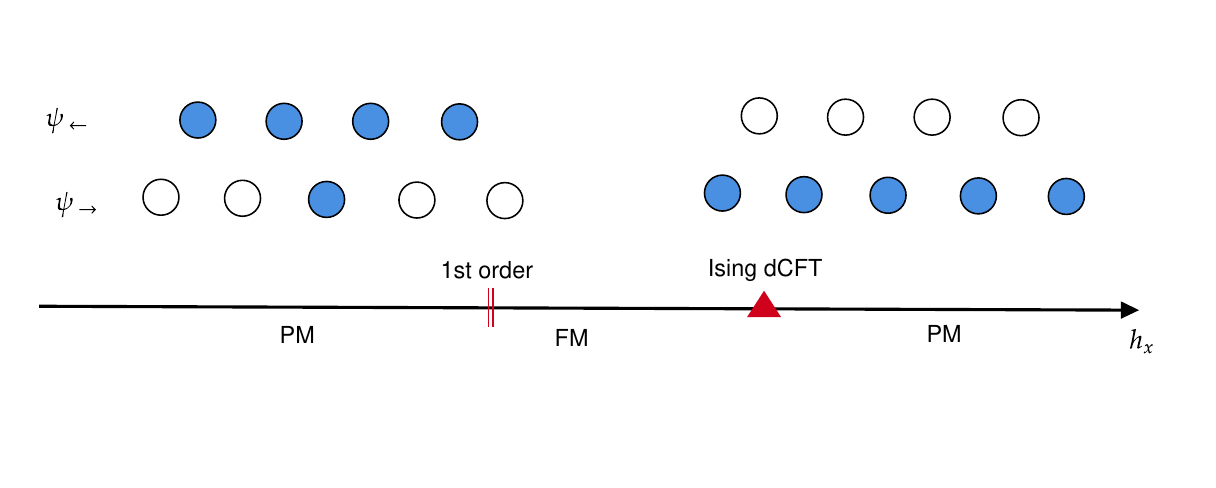}
    \caption{Schematic phase diagram of the Ising model with monodromy defect on the fuzzy sphere.}
    \label{fig:phase_IS}
\end{figure}

In the ordinary Ising model, the transformation $\prod_i Z_i$ relates $h_x$ and $-h_x$, and both limits $h_x\rightarrow\pm \infty$ are paramagnetic. In the twisted model, however, this transformation exchanges $\psi_\rightarrow$ and $\psi_\leftarrow$, which obey different monodromy conditions and have different numbers of orbitals, so the two $h_x\rightarrow\pm \infty$ limits corresponds to different phases. We take $N_e=N=2s+1$. The $h_x\rightarrow\infty$ phase has fully filled $\psi_\rightarrow$ orbitals, with $L_z=0,R_{yz} =(-1)^{\frac{N(N-1)}{2}}$. In the $h_x\rightarrow-\infty$ limit, all $\psi_\leftarrow$ orbitals are occupied, and one additional electron occupies a $\psi_\rightarrow$ orbital, this state sits at integer $L_z$ for odd $N$, and half-integer $L_z$ for even $N$. We expect there is a first order phase transition where the two competing ground-state energies cross, a sketch of phase diagram is shown in figure \ref{fig:phase_IS}.

\begin{figure}[t]
    \centering
    \begin{minipage}[t]{0.45\linewidth}
    \stackinset{l}{2mm}{t}{2mm}{\textbf{a.}}{
        \includegraphics[width=\linewidth]{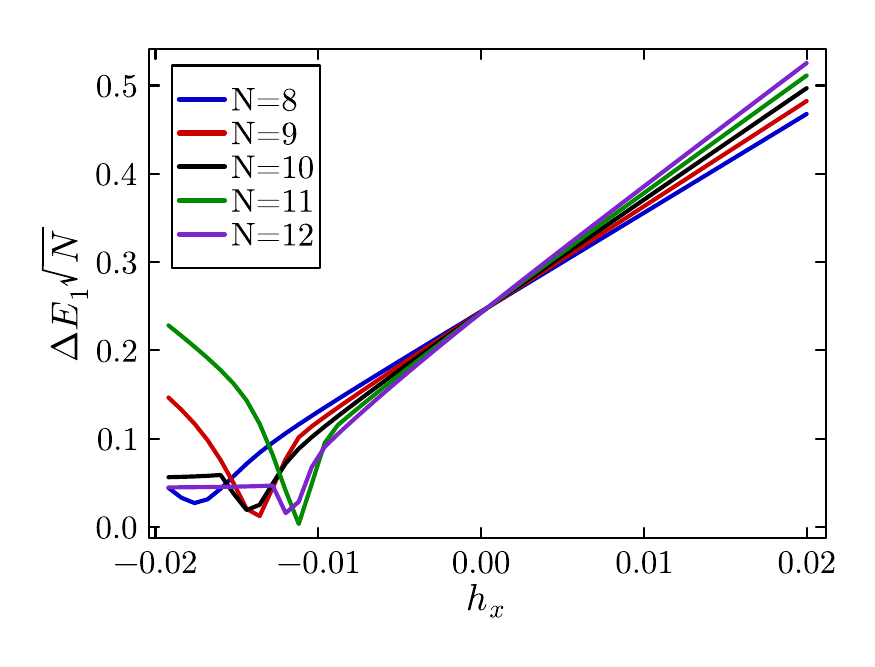}
        }
    \end{minipage}
    \begin{minipage}[t]{0.45\linewidth}
    \stackinset{l}{2mm}{t}{2mm}{\textbf{b.}}{
        \includegraphics[width=\linewidth]{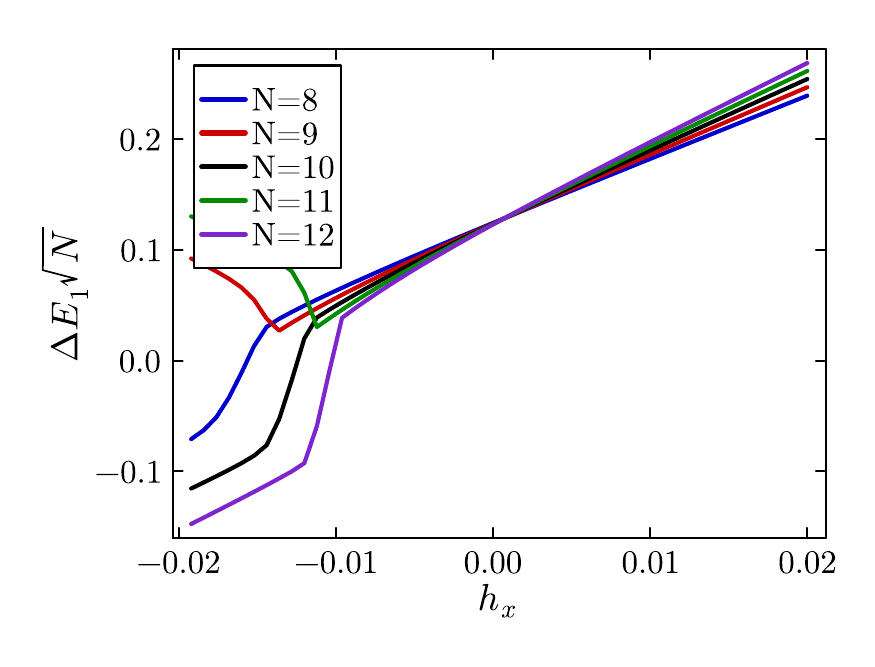}
        }
    \end{minipage}
    \caption{Rescaled energy gaps $\Delta E\sqrt{N}$ in the (a) $(0^+)$ and (b) $(1/2)$ sectors.}
    \label{fig: IS_fs_gap}
\end{figure}

In absence of a uniform order parameter \footnote{The order-parameter field $\sigma$ has twisted boundary condition in the symmetry-broken phase, so the usual scaling law $\langle M^2\rangle\simeq R^4$ does not apply.}, we compute the first energy gap in the $(0^+)$ sector and the first energy gap in the $(\frac{1}{2})$ sector for different values of $N$ to locate the critical point, see figure \ref{fig: IS_fs_gap}. There is a first order phase transition near $h_x\simeq -0.01$, characterized by a level crossing in the neutral $(0^+)$ sector for odd $N$, while for even $N$ a lower energy state appears in the $(\frac{1}{2})$ sector.

In addition, a second order phase transition appears at $h_x\simeq0.00$, where the energy gaps scale as $\Delta E\simeq \frac{1}{\sqrt{N}} $ in both odd and even sectors.

\begin{figure}[t]
    \centering
    \includegraphics[width=\linewidth]{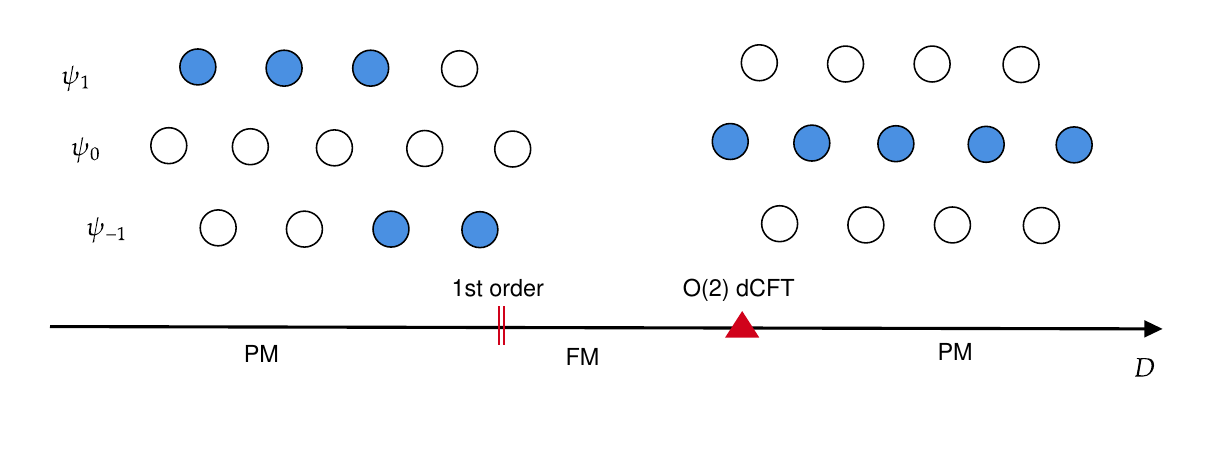}
    \caption{Schematic phase diagram of the $O(2)$ model with monodromy defect on the fuzzy sphere.}
    \label{fig:phase_WF}
\end{figure}

For the $O(2)$ model, the $D\rightarrow+\infty$ limit gives a paramagnetic phase with fully filled $\psi_0$ orbitals, which has $\mathbb Z_{2,RC}$ quantum number $(-1)^{\frac{N(N+1)}{2}}$. At small $D$, the dominant interaction is the in-plane spin interaction, which favors an $O(2)$-breaking ferromagnetic phase. At $D\rightarrow-\infty$ limit $\psi_0$ orbitals are unoccupied, for even $N$ the ground state is in the $(L_z,Q)=(0,0)$ sector. For odd $N$, since $\psi_1$ and $\psi_{-1}$ have unequal occupation numbers, the ground state develops a non-zero $L_z$ quantum number. A first order phase transition is expected at intermediate $D$, a sketch of the phase diagram is shown in figure \ref{fig:phase_WF}.

\begin{figure}[t]
    \centering
    \begin{minipage}[t]{0.45\linewidth}
    \stackinset{l}{2mm}{t}{2mm}{\textbf{a.}}{
        \includegraphics[width=\linewidth]{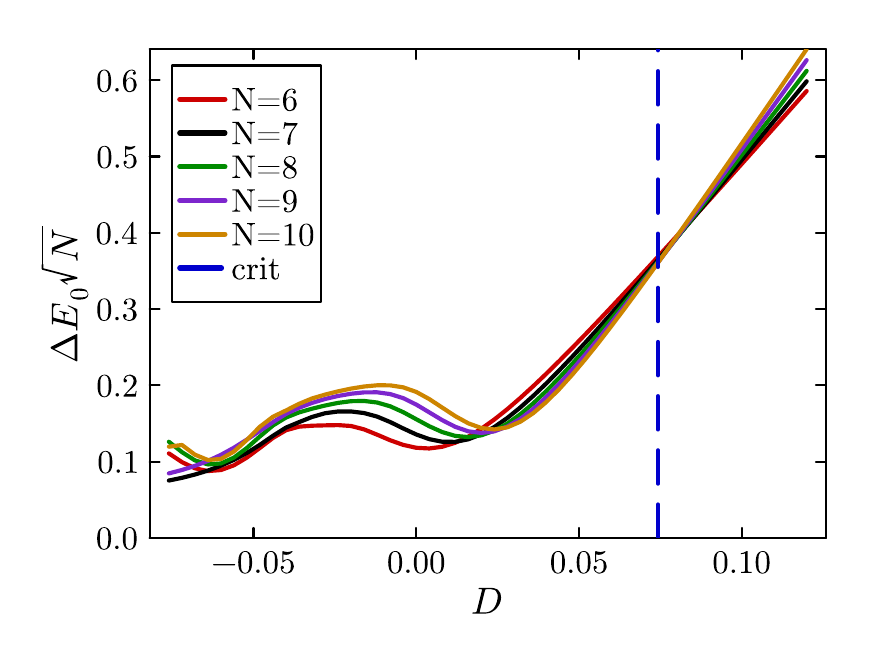}
        }
    \end{minipage}
    \begin{minipage}[t]{0.45\linewidth}
    \stackinset{l}{2mm}{t}{2mm}{\textbf{b.}}{
        \includegraphics[width=\linewidth]{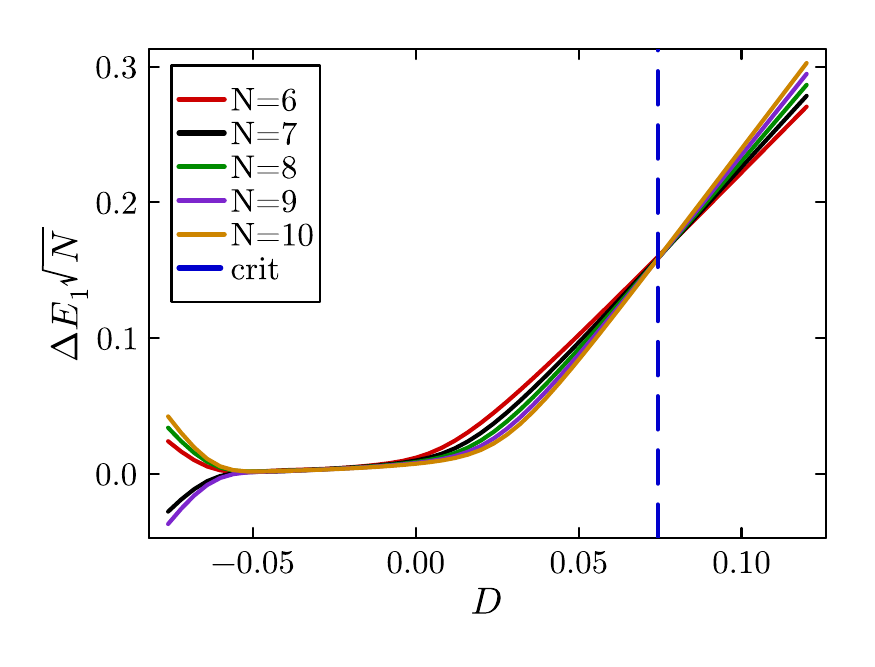}
        }
    \end{minipage}
    \caption{Rescaled (a) neutral gap and (b) charge-1 gap as a function of $D$ at $\nu = 1/4$. The coupling $U_1, V_1$ are chosen to be the tuned values.}
    \label{fig: O2_fs_gap}
\end{figure}

Numerical computation at $N=6,7,...,10$, at $D\simeq -0.06$ reveals a lower-energy state in $(L_z,Q)=(\nu,1)$ for odd $N$, see figure \ref{fig: O2_fs_gap}(a,b), indicating a first order phase transition. In addition, at $D_c\simeq 0.05$ we find a crossing point of $\Delta E\sqrt{N}$ for both the neutral gap and the $Q=1$ gap, which we identify with the $O(2)$ dCFT critical point.

\section{Relevant perturbations at large sizes}\label{app: rel_perturbation}
\begin{figure}[t]
    \centering
    \begin{minipage}[t]{0.45\linewidth}
    \stackinset{l}{2mm}{t}{2mm}{\textbf{a.}}{
        \includegraphics[width=\linewidth]{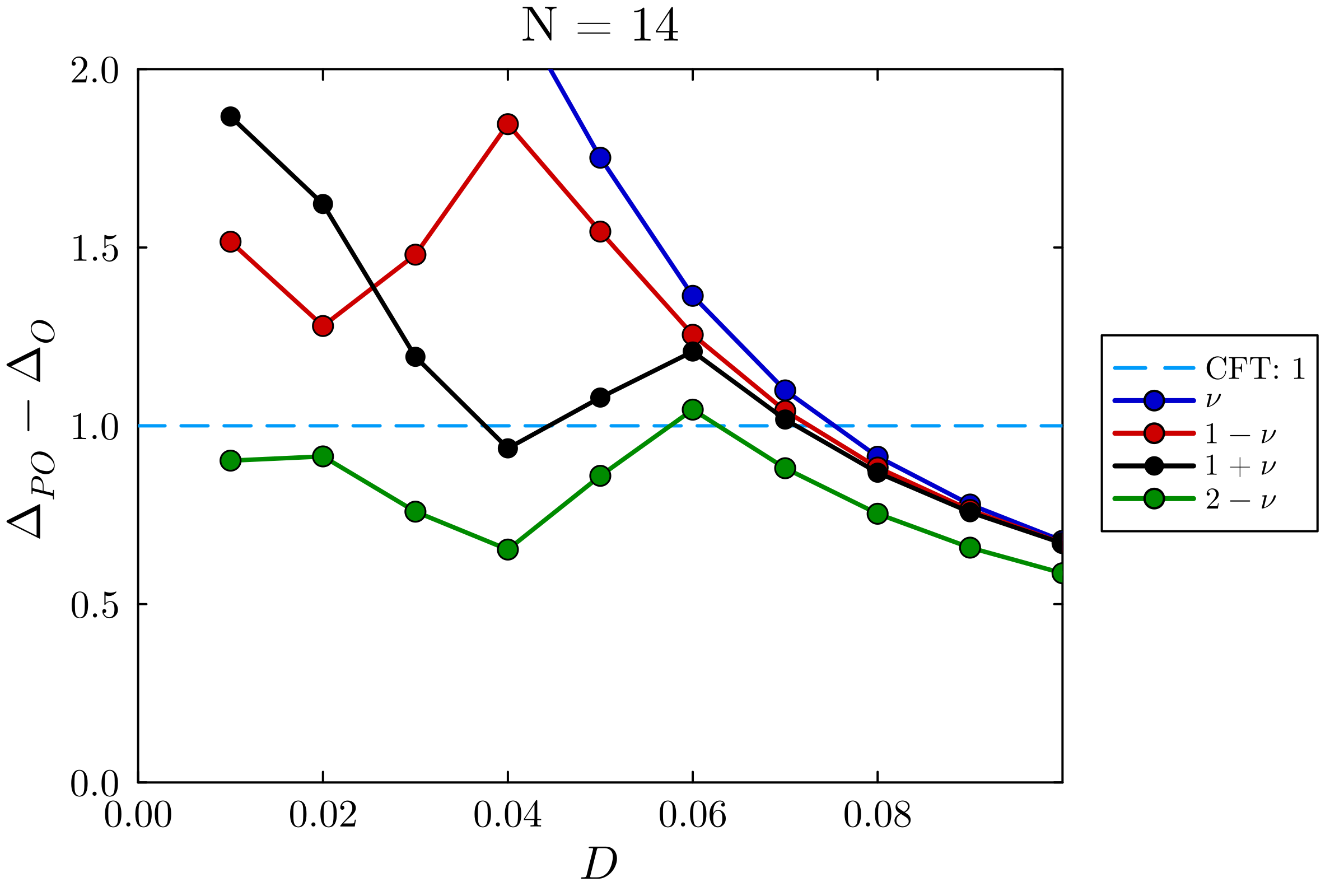}
        }
    \end{minipage}
    \begin{minipage}[t]{0.45\linewidth}
    \stackinset{l}{2mm}{t}{2mm}{\textbf{b.}}{
    \includegraphics[width=\linewidth]{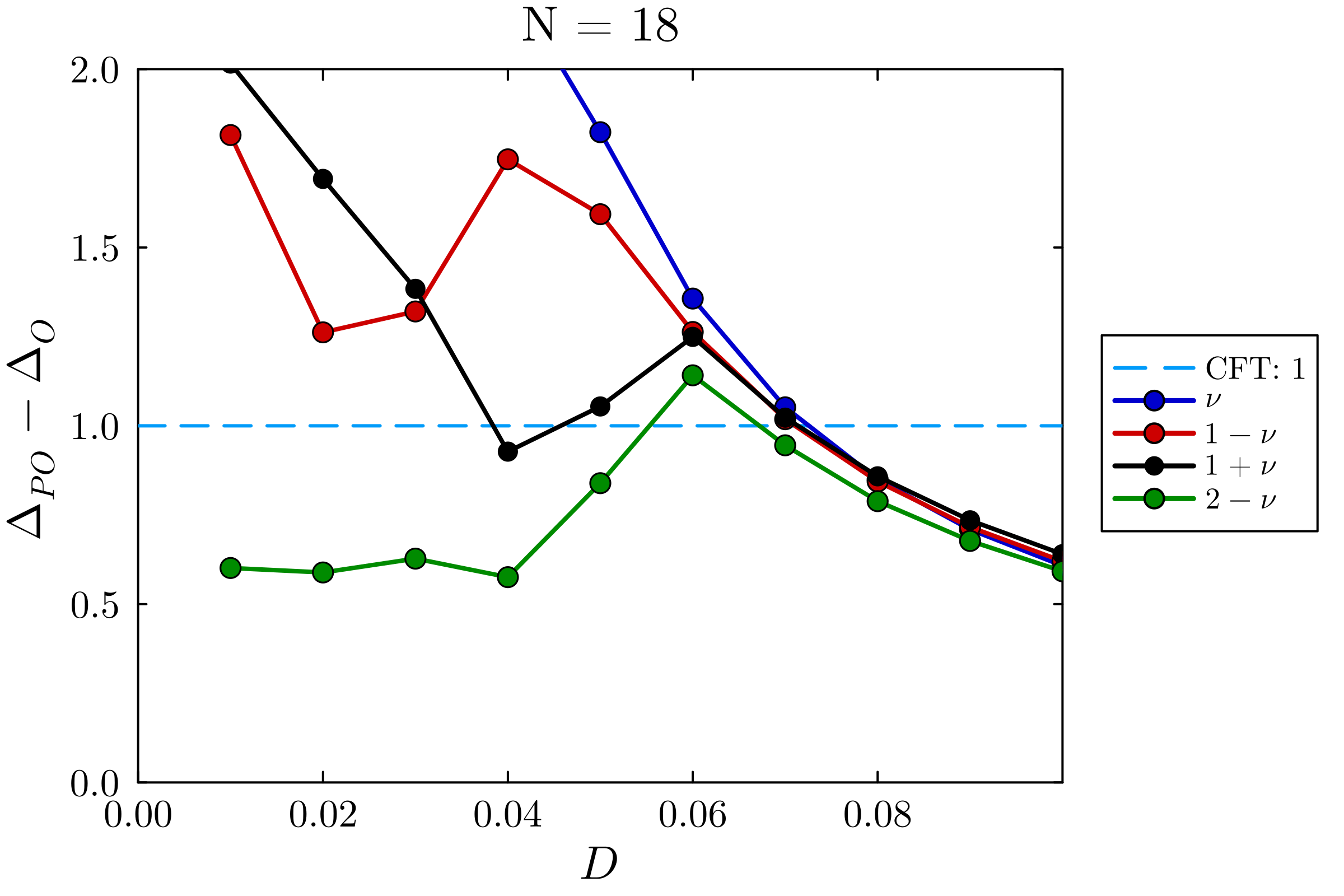}
        }
    \end{minipage}
    \begin{minipage}[t]{0.45\linewidth}
    \stackinset{l}{2mm}{t}{2mm}{\textbf{c.}}{
        \includegraphics[width=\linewidth]{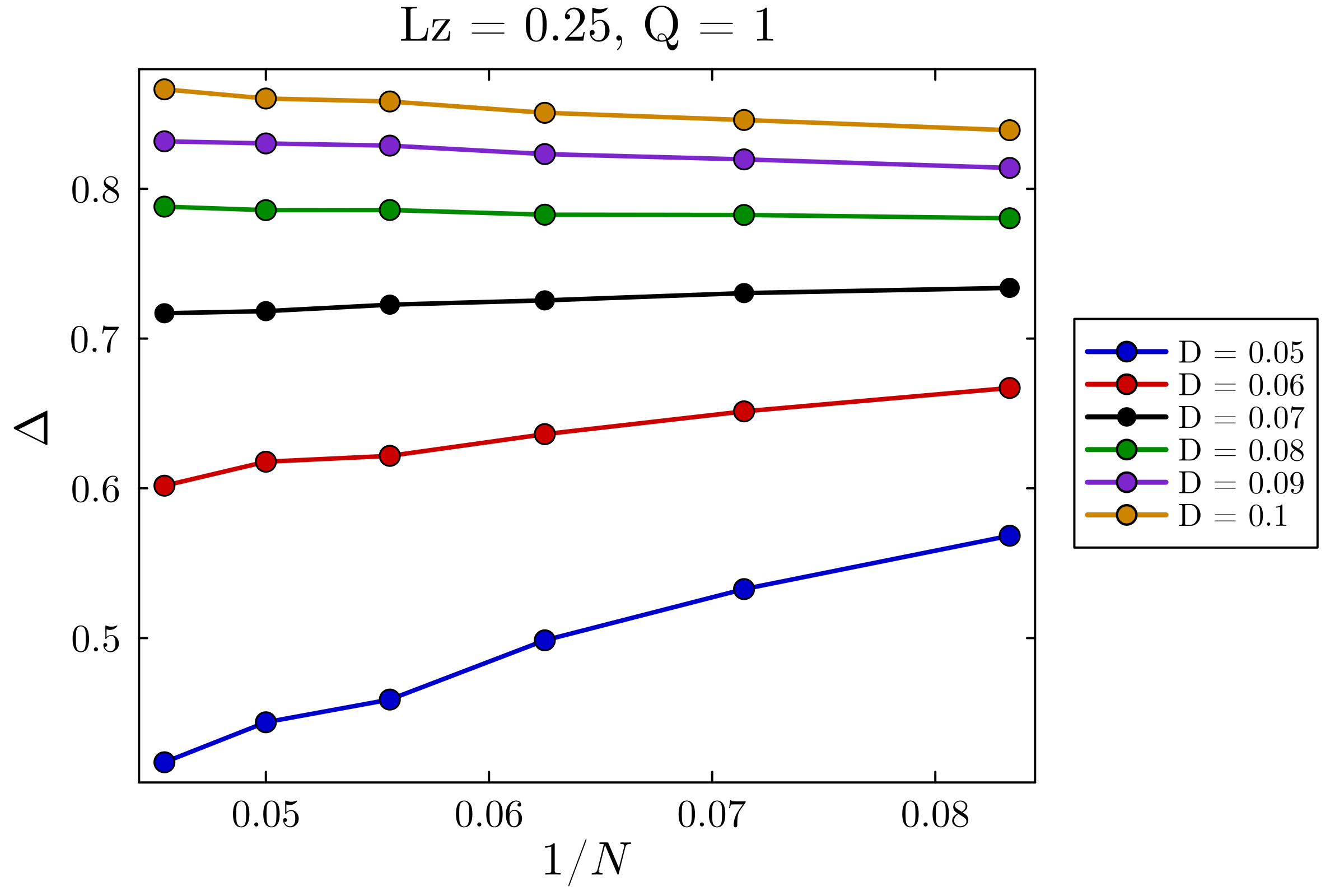}
        }
    \end{minipage}
    \begin{minipage}[t]{0.45\linewidth}
    \stackinset{l}{2mm}{t}{2mm}{\textbf{d.}}{
        \includegraphics[width=\linewidth]{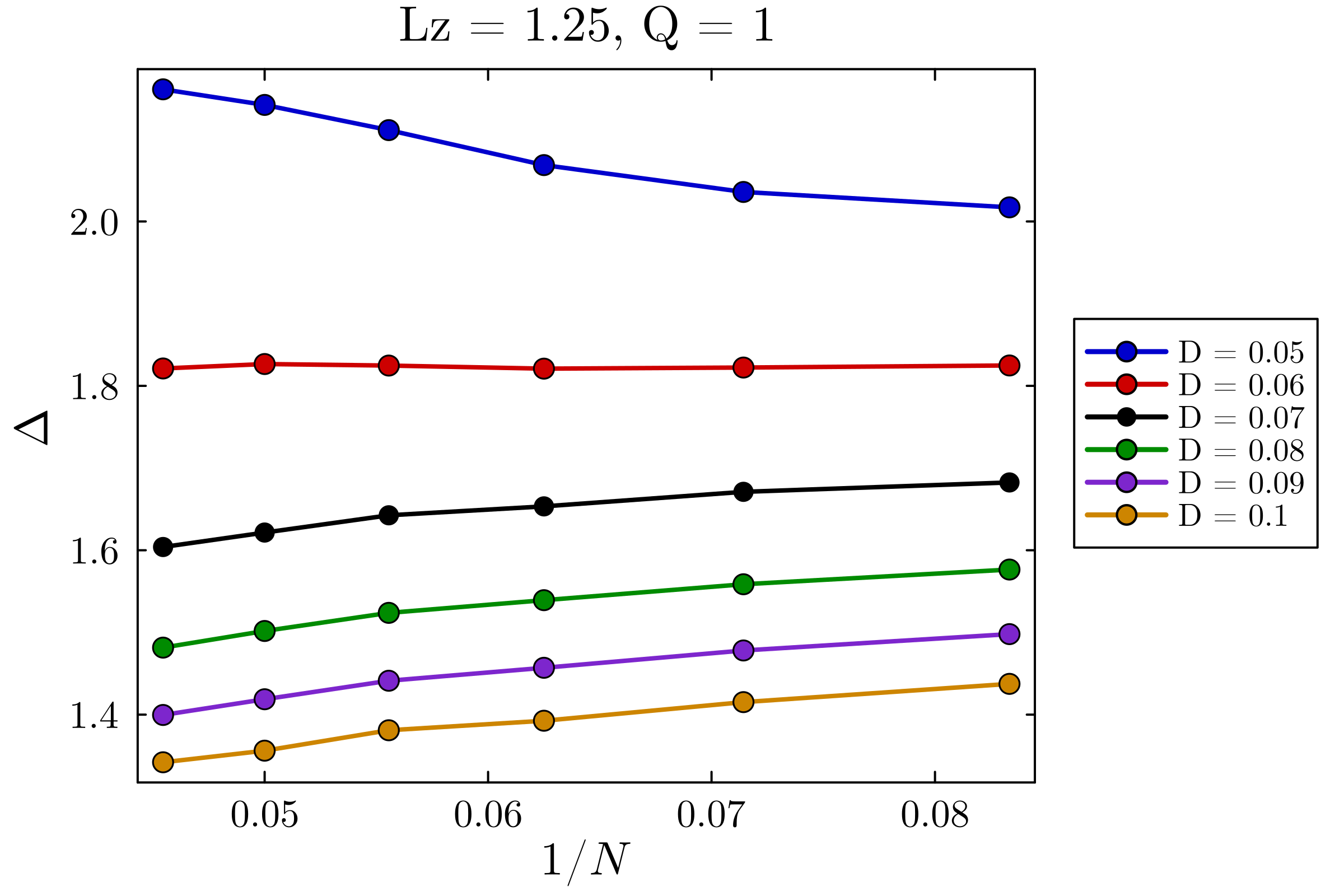}
        }
    \end{minipage}
    \caption{(a) Conformal level spacing $\Delta_{PO}-\Delta_{O}$, several sectors at (a) $N=14$, (b) $N=18$, where $O$ is a $SO(2,1)$ primary. (c,d) Finite-size dependence of primary scaling dimension of $\Psi_\nu$ and $\Psi_{1+\nu}$ respectively, as a function of $D$. All panels show results for $\nu=1/4$.}
    \label{fig: D_perturbation}
\end{figure}

Relevant perturbations may become dominant at large system size. We illustrate this effect by varying $D$. Since $\Psi^\dagger (S^z)^2 \Psi$ is a neutral scalar, shifting $D\rightarrow D_c+\delta D$ induces a coupling to the relevant bulk operator $S(x)$, driving a RG flow away from the CFT fixed point,
\begin{equation}
    S_{CFT} + \alpha(D-D_c)\int d^3 x S(x).
\end{equation}
To be specific, the scaling dimension has finite-size scaling in the perturbative regime:
\begin{equation}
    \Delta(D,N) = f_0N^{(3-\Delta_S)/2}(D-D_c)+ \Delta(\infty) + a_1 N^{-\omega/2}+.... 
\end{equation}
For $\nu = 1/4$, imposing the conformal level-spacing condition $\Delta_{PO}-\Delta_O=1$ in several sectors gives $D_c\simeq 0.070(5)$. We compare the spectra at $N=12,14,...,22$ for the nearby values $D= D_c +\delta D$, and see that a small but non-zero $\delta D$ drives growing deviations at large $N$, see figure \ref{fig: D_perturbation}(c-d). The proximity of the descendant-primary spacings to one therefore provides a diagnostic of how accurately the critical point has been located.

\begin{figure}[t]
    \centering
     \begin{minipage}[t]{0.9\linewidth}
     \includegraphics[width=\linewidth]{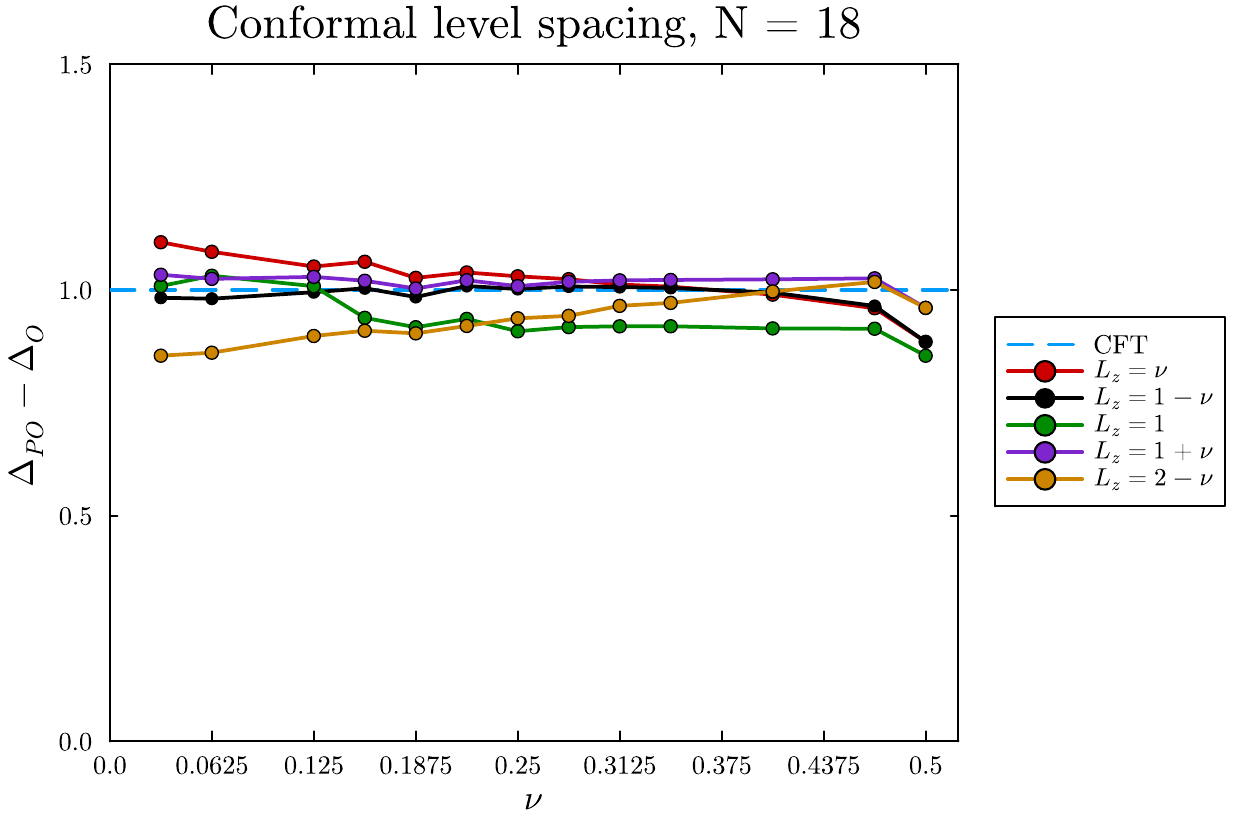}  
    \end{minipage}
    \caption{Conformal level spacing $\Delta_{PO}-\Delta_O$ at $N=18$ for sampled value of $\nu$, where $O$ is the lowest primary in each sector.}
    \label{fig:DMRG_level spacing}
\end{figure}

Across the sampled values of $\nu$, optimizing $P(U_1,V_1,D,\nu)$ using $N=7,...,11$ leaves a deviation of $\Delta_{PO}-\Delta_O$ from one by order $0.2$ at $N=18$. These may reflect the finite-size drift of estimated critical parameters. We introduce a second tuning step, adjusting $D$ to minimize $|\Delta_{PO}-\Delta_O-1|$ at $N=18$ for several primaries, the resulting level spacings are shown in figure \ref{fig:DMRG_level spacing}. The DMRG calculations use a maximum bond dimension of $\chi=5000$ and a truncation cutoff of $10^{-8}$. Table \ref{tab:dmrg_N18} lists the scaling dimensions of the lowest operators at these tuned parameters. 

\begin{table}[t]
\centering
\renewcommand{\arraystretch}{1.2}
\begin{tabular}{c|ccccccc}
\hline\hline
$\nu$ & $\Delta_{\hat S}$ & $\Delta_{\nu}$ & $\Delta_{1-\nu}$ & $\Delta_{1}$ & $\Delta_{1+\nu}$ & $\Delta_{2-\nu}$ & $\Delta_{2}$ \\
\hline
$0.03125$ & 1.485373 & 0.568266 & 1.363433 & 2.000000 & 1.471314 & 2.394764 & 2.830842 \\
$0.06250$ & 1.502381 & 0.587196 & 1.311557 & 2.000000 & 1.479302 & 2.330387 & 2.814595 \\
$0.12500$ & 1.555039 & 0.637627 & 1.227738 & 2.000000 & 1.513976 & 2.243424 & 2.769346 \\
$0.15625$ & 1.597236 & 0.659124 & 1.213048 & 2.000000 & 1.560304 & 2.230877 & 2.807001 \\
$0.18750$ & 1.632799 & 0.684133 & 1.162951 & 2.000000 & 1.563734 & 2.161187 & 2.757516 \\
$0.21875$ & 1.661550 & 0.715817 & 1.152923 & 2.000000 & 1.613382 & 2.162231 & 2.804267 \\
$0.25000$ & 1.690985 & 0.730907 & 1.116980 & 2.000000 & 1.631209 & 2.117529 & 2.786200 \\
$0.28125$ & 1.734662 & 0.768692 & 1.099691 & 2.000000 & 1.668367 & 2.098197 & 2.813878 \\
$0.31250$ & 1.766864 & 0.801701 & 1.077264 & 2.000000 & 1.697864 & 2.064840 & 2.806396 \\
$0.34375$ & 1.799743 & 0.826170 & 1.053215 & 2.000000 & 1.726533 & 2.034004 & 2.825362 \\
$0.40625$ & 1.868524 & 0.884169 & 1.013706 & 2.000000 & 1.777298 & 1.961849 & 2.839812 \\
$0.46875$ & 1.933532 & 0.945856 & 0.984854 & 2.000000 & 1.817819 & 1.879324 & 2.834295 \\
$0.50000$ & 1.959924 & 0.976209 & 0.976282 & 2.000000 & 1.784441 & 1.784326 & 2.772014 \\
\hline\hline
\end{tabular}
\caption{Defect scaling dimensions obtained by DMRG at $N=18$, using parameters obtained by the two-step tuning.}
\label{tab:dmrg_N18}
\end{table}

\begin{table}[t]
\centering
\setlength{\tabcolsep}{18pt}
\renewcommand{\arraystretch}{1.2}
\begin{tabular}{c|cccc}
\hline\hline
$\nu$ & $b_{S \hat S}/b_{S \hat D}$ & err. & $b_{\phi  \Psi_{\nu-1}}/b_{\phi  \Psi_{\nu-2}}$ & err. \\
\hline
0.03125 & 5.75403 & 3.16e-01 & 1.10654 & 6.02e-04 \\
0.0625 & 4.31148 & 7.08e-02 & 1.10813 & 3.67e-04 \\
0.125 & 2.33462 & 4.79e-03 & 1.10690 & 5.53e-04 \\
0.15625 & 1.95105 & 8.93e-03 & 1.10928 & 6.25e-04 \\
0.1875 & 1.70920 & 8.26e-03 & 1.11079 & 7.74e-04 \\
0.21875 & 1.38562 & 1.44e-03 & 1.10289 & 1.08e-03 \\
0.25 & 1.30258 & 8.62e-03 & 1.11219 & 7.26e-04 \\
0.28125 & 1.12705 & 3.46e-03 & 1.10308 & 1.13e-03 \\
0.3125 & 1.01397 & 6.20e-03 & 1.09311 & 1.10e-03 \\
0.34375 & 0.88285 & 8.81e-03 & 1.09292 & 1.10e-03 \\
0.40625 & 0.75216 & 1.65e-02 & 1.07548 & 1.29e-03 \\
0.46875 & 0.36138 & 4.32e-03 & 1.03414 & 1.14e-03 \\
0.5 & 0.11458 & 9.29e-03 & 0.99514 & 1.04e-03 \\
\hline\hline
\end{tabular}
\caption{Ratios of DOE coefficients obtained from fits of form $a+b/N$ using $N=12,...,22$, for sampled values of $\nu$. The value errors are extrapolation errors.}
\label{tab:ratio_fit}
\end{table}

\section{Conformal cost functions}\label{app: cost functions}
The cost function $Q(N)$ measures deviations of descendant-primary level spacings from the conformal value. For the Ising monodromy dCFT, let $E_{O}$ denote the energy gap associated with $O$, normalized so that $E_D=2$. We define:
\begin{equation}
\begin{split}
    Q(N) =& \frac{\sum _i W_i(\delta\Delta_i-\delta\Delta_i^{CFT})^2}{\sum _i W_i},\\
     \delta\Delta=& [E_{PS}- E_{S}, E_{P\psi_{1/2}}- E_{\psi_{1/2}},E_{P\psi_{3/2}}- E_{\psi_{3/2}}, E_{PD}- 2],\\
     \delta\Delta^{CFT}=& [1,1,1,1].\\
        W_i = & 1/\Delta_i \\
      \Delta=& [E_{PS}, E_{P\psi_{1/2}},E_{P\psi_{3/2}},  E_{PD}].
\end{split}
\end{equation}
For the $O(2)$ monodromy dCFT, we use:
\begin{equation}
\begin{split}
    Q(N) =& \frac{\sum _i W_i(\delta\Delta_i-\delta\Delta_i^{CFT})^2}{\sum _i W_i},\\
     \delta\Delta=& [E_{P\Psi_\nu}- E_{\Psi_\nu}, E_{P\bar\Psi_{1-\nu}}- E_{\bar\Psi_{1-\nu}},E_{P\Psi_{1+\nu}}- E_{\Psi_{1+\nu}}, E_{P\bar\Psi_{2-\nu}}- E_{\bar\Psi_{2-\nu}}],\\
     \delta\Delta^{CFT}=& [1,1,1,1],\\
     W_i = & 1/\Delta_i ,\\
      \Delta=& [E_{P\Psi_\nu}, E_{P\bar \Psi_{1-\nu}},E_{P\Psi_{1+\nu}}, E_{P\bar\Psi_{2-\nu}}].
\end{split}
\end{equation}

\newpage
\bibliographystyle{JHEP}
\bibliography{biblio}
\end{document}